\documentclass[
 preprint,
 superscriptaddress,
 amsmath,amssymb,
 aps,
 prl,
 longbibliography,
 showpacs
]{revtex4-1}
\usepackage{graphicx}
\usepackage{dcolumn}
\usepackage{bm}
\usepackage{epstopdf}
\begin{document}
\title{Cavity Quantum Electrodynamics with Second-Order Topological Corner State}
\author{Xin Xie}
\thanks{Contributed equally to this work.}
\affiliation{Beijing National Laboratory for Condensed Matter Physics, Institute of Physics, Chinese Academy of Sciences, Beijing 100190, China}
\affiliation{CAS Center for Excellence in Topological Quantum Computation and School of Physical Sciences, University of Chinese Academy of Sciences, Beijing 100049, China}
\author{Weixuan Zhang}
\thanks{Contributed equally to this work.}
\affiliation{Key Laboratory of advanced optoelectronic quantum architecture and measurements of Ministry of Education, School of Physics, Beijing Institute of Technology, 100081, Beijing, China}
\affiliation{Beijing Key Laboratory of Nanophotonics $\&$ Ultrafine Optoelectronic Systems, Micro-nano Center, School of Physics, Beijing Institute of Technology, 100081, Beijing, China}
\author{Xiaowu He}
\affiliation{State Key Laboratory of Superlattices and Microstructures, Institute of Semiconductors Chinese Academy of Sciences, Beijing 100083, China}
\author{Shiyao Wu}
\author{Jianchen Dang}
\author{Kai Peng}
\author{Feilong Song}
\author{Longlong Yang}
\affiliation{Beijing National Laboratory for Condensed Matter Physics, Institute of Physics, Chinese Academy of Sciences, Beijing 100190, China}
\affiliation{CAS Center for Excellence in Topological Quantum Computation and School of Physical Sciences, University of Chinese Academy of Sciences, Beijing 100049, China}
\author{Haiqiao Ni}
\affiliation{State Key Laboratory of Superlattices and Microstructures, Institute of Semiconductors Chinese Academy of Sciences, Beijing 100083, China}
\author{Zhichuan Niu}
\affiliation{State Key Laboratory of Superlattices and Microstructures, Institute of Semiconductors Chinese Academy of Sciences, Beijing 100083, China}
\author{Can Wang}
\author{Kuijuan Jin}
\affiliation{Beijing National Laboratory for Condensed Matter Physics, Institute of Physics, Chinese Academy of Sciences, Beijing 100190, China}
\affiliation{CAS Center for Excellence in Topological Quantum Computation and School of Physical Sciences, University of Chinese Academy of Sciences, Beijing 100049, China}
\affiliation{Songshan Lake Materials Laboratory, Dongguan, Guangdong 523808, China}
\author{Xiangdong Zhang}
\email{zhangxd@bit.edu.cn}
\affiliation{Key Laboratory of advanced optoelectronic quantum architecture and measurements of Ministry of Education, School of Physics, Beijing Institute of Technology, 100081, Beijing, China}
\affiliation{Beijing Key Laboratory of Nanophotonics $\&$ Ultrafine Optoelectronic Systems, Micro-nano Center, School of Physics, Beijing Institute of Technology, 100081, Beijing, China}
\author{Xiulai Xu}
\email{xlxu@iphy.ac.cn}
\affiliation{Beijing National Laboratory for Condensed Matter Physics, Institute of Physics, Chinese Academy of Sciences, Beijing 100190, China}
\affiliation{CAS Center for Excellence in Topological Quantum Computation and School of Physical Sciences, University of Chinese Academy of Sciences, Beijing 100049, China}
\affiliation{Songshan Lake Materials Laboratory, Dongguan, Guangdong 523808, China}

\begin{abstract}

Topological photonics provides a new paradigm in studying cavity quantum electrodynamics with robustness to disorder. In this work, we demonstrate the coupling between single quantum dots and the second-order topological corner state. Based on the second-order topological corner state, a topological photonic crystal cavity is designed and fabricated into GaAs slabs with quantum dots embedded. The coexistence of corner state and edge state with high quality factor close to 2000 is observed. The enhancement of photoluminescence intensity and emission rate are both observed when the quantum dot is on resonance with the corner state. This result enables the application of topology into cavity quantum electrodynamics, offering an approach to topological devices for quantum information processing.

\end{abstract}
\maketitle



\section{Introduction}

Cavity quantum electrodynamics (CQED) studies the interaction between photons and quantum emitters, including the strong and weak coupling regime, which has widespread applications \cite{vahala2003optical,khitrova2006vacuum,lodahl2015interfacing}. For instance, CQED systems are widely proposed for the realization of quantum information processing \cite{imamog1999quantum}. Especially, the weak coupling can greatly enhance the spontaneous emission rate of quantum emitters \cite{purcell1995spontaneous,englund2005controlling,Laucht_2009}, which can be used to optimize the photonic devices, such as high-efficiency single photon source \cite{liu2018high,Chang2019Efficient,strauf2007high}, ultra-fast qubit gate \cite{carter2013quantum,sweeney2014cavity} and low-threshold laser \cite{lonvcar2002low}. Up to now, solid-state CQED experiments have been implemented in a wide range of photonic nanocavities, including whispering gallery modes \cite{srinivasan2007linear}, Anderson-localized modes \cite{sapienza2010cavity}, and photonic crystal (PhC) cavities \cite{yoshie2004vacuum,hennessy2007quantum,brossard2010strongly,qian2018two,qian2019enhanced} with coupled two-level systems. These photonic nanocavities are optimized for high quality factor (Q) and small mode volume to enhance the coupling strength, and they are always strongly affected by defects and disorders introduced by fabrication imperfections, environment perturbations, etc. The emerging field of topological photonics provides a new paradigm to solve the problem, offering an approach to the development of photonic devices with robustness to defects and disorders.

So far, the application of topology in optics has been investigated in many areas \cite{khanikaev2017two,ozawa2019topological,lu2014topological,wu2015scheme,blanco2016topological,rechtsman2013photonic,rechtsman2013topological,haldane2008possible,wang2009observation,wang2008reflection,
fang2012realizing,hafezi2011robust,khanikaev2013photonic,hafezi2013imaging,harari2018topological,bandres2018topological,zhao2018topological,parto2018edge,ota2018topological,
bahari2017nonreciprocal,st2017lasing,ota2019active,chen2014experimental,smirnova2019nonlinear}, such as one-way waveguide \cite{wang2009observation,wang2008reflection,haldane2008possible,fang2012realizing,hafezi2011robust,khanikaev2013photonic,hafezi2013imaging}, topological lasers \cite{harari2018topological,bandres2018topological,zhao2018topological,parto2018edge,ota2018topological,bahari2017nonreciprocal,st2017lasing,ota2019active}, which are mainly in the classical domain. The combination of topology with quantum regime will bring more interesting phenomena and physics \cite{blanco2018topological,tambasco2018quantum,mittal2018topological,wang2019topological,wang2019direct}. Especially, the coupling between quantum emitters and topologically protected state will exhibit robust strong light-matter interaction, enabling a topological quantum optics interface \cite{barik2018topological,mehrabad2019chiral}. For example, the coupling to topological edge state enables the robust chiral emission of quantum dots (QDs) \cite{barik2018topological}. More importantly, the coupling to topological nanocavity will have more widespread applications in development of photonic devices and quantum optics devices for quantum information with built-in protection, which has not been demonstrated. To investigate CQED with topological state, both high-quality topological nanocavity and a good matching with quantum emitters are required. Recently, a new class of higher-order topological insulators have been proposed and experimentally demonstrated in different systems \cite{xie2018second,xie2019visualization,ota2019photonic,chen2019direct,langbehn2017reflection,ezawa2018higher,ni2019observation,noh2018topological,benalcazar2017quantized,
benalcazar2017electric,imhof2018topolectrical,peterson2018quantized,serra2018observation,mittal2019photonic,xue2019acoustic, zhang2019second,schindler2018higher,el2019corner,zhang2020topolectrical,bao2019topoelectrical,xue2019quantized,ni2019demonstration,zhang2020symmetry}, including 2D PhC slab where 0D topological corner state has been observed \cite{ota2019photonic,chen2019direct}. This high-order topological state provides an ideal platform to design topological nanocavity for the investigation of CQED.

In this work, we report on the coupling between the second-order topological corner state and single QDs. Based on the generalised 2D Su-Schrieffer-Heeger (SSH) model, a topological PhC cavity is designed and fabricated into GaAs slab with QDs embedded. The Q is optimized by shifting the non-trivial PhC away from the corner. The existence of topological corner and edge states with high Q are observed by photoluminescence (PL) spectra. The emission intensity of single QD coupled to the corner state is enhanced by a factor of about 4, and the emission rate is enhanced by a factor of 1.3. Our results demonstrate the potential of the application of topology into CQED, enabling the development of quantum information processing.

\section{Design and optimization of topological corner state}

\begin{figure}
\centering
\includegraphics[scale=0.43]{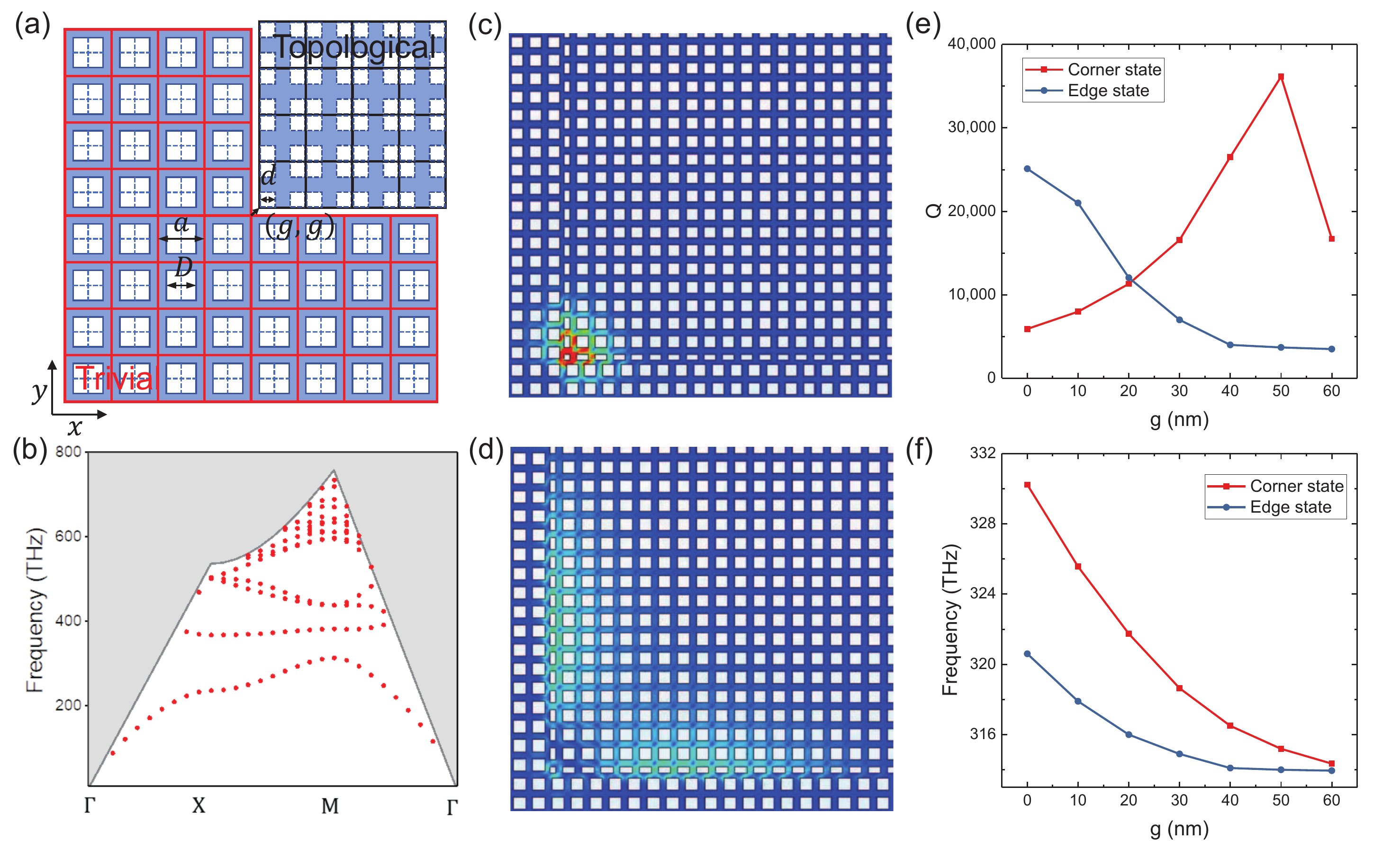}
\caption{(a) Schematic of the 2D topological PhC cavity in a square shape with square air holes with length of $D$ ($d=D/2$). Unit cells with red (black) outline correspond to trivial (topological) PhC. The topological PhC is shifted away from the corner by $(g,g)$ to optimize Q of the corner state. (b) Bulk band structure of the 2D PhC slab with $a=$ 280 nm, $D=0.64a$ and $t=$ 150 nm. Electric field profile of (c) topological corner state at 321.11 THz and (d) topological edge state at 314.8 THz with $g=$ 20 nm. (e) Q and (f) frequency of corner state and edge state with different $g$. }
\label{f1}
\end{figure}

Previous investigations have shown that the 0D corner state can exist in the 2D topological insulators protected by the quantized bulk quadrupole polarization \cite{benalcazar2017electric,benalcazar2017quantized,imhof2018topolectrical,peterson2018quantized,serra2018observation,mittal2019photonic} or quantized edge dipole polarization \cite{xie2018second,xie2019visualization,ota2019photonic,chen2019direct}, which is related to the 2D Zak phase, $\theta^{Zak}=(\theta_x,\theta_y)$ defined as
\begin{eqnarray}
\theta_i&=&\int\mathrm{d}k_{x}\mathrm{d}k_{y}Tr[A_{i}(k_{x},k_{y})],
\label{eq.1}
\end{eqnarray}
where $i=x$ or $y$ and $A_{i}(k_{x},k_{y})$ is the Berry connection \cite{zak1989berry}. Many recent works \cite{ota2019photonic,chen2019direct} have shown that the Wannier-type 0D corner state, which is induced by the non-trivial edge dipole polarization, can be easily realized in the optical system by combing two different photonic crystals with distinct topologies but identical band gap. In this case, we design a topological nanocavity in the 2D PhC slab based on such 0D corner state, as shown in Fig. \ref{f1}(a). The cavity consists of two topologically distinct PhC slabs in square shape with the same period $a$ and thickness of $t$, which are distinguished by different colors of the outline. It is clearly shown that the four sub-squares in unit cell with the black (red) outline are far away from (adjacent to) each other, corresponding to a topological (trivial) phase with the 2D Zak phase being $\theta^{Zak}=(\pi,\pi)$ ($(0,0)$) \cite{ota2019photonic,chen2019direct}. Additionally, due to the same period and size of the structures, these two PhC slabs share the common band structure, as presented in Fig. \ref{f1}(b). By suitably combing the two topologically distinct PhC slabs, the quantized edge dipole polarizations along $x$- and $y$-axis can induce the 0D corner state in the band gap, and the related edge states get opened. Fig. \ref{f1}(c) and (d) show the electric field profiles of the corner and edge states. Contrast to the dispersive distribution of 1D edge state, the corner state is tightly localized at the intersection of two types of PhC slab, which has a much smaller mode volume.

In addition to a small mode volume, the high Q is also important for the coupling between quantum emitters and nanocavity. The Q of corner state is optimized by slightly shifting the topological PhC away from the corner by $(g,g)$, as shown in Fig. \ref{f1}(a). In that case, Q of corner state first increases and then decreases with increasing $g$, while Q of edge states decrease monotonically, as shown in Fig. \ref{f1}(e). When $g=$ 20 nm, the Q of corner state and one of edge states are comparable with the magnitude of $10^4$. However, their mode volumes show a big difference, about $0.23(\lambda/n)^{3}$ for the corner state and $6.59(\lambda/n)^{3}$ for the edge state, where $\lambda$ is the resonant wavelength of the cavity and $n$ is the refractive index of GaAs. Meanwhile, the corner state and edge state show redshift with increasing $g$, as shown in Fig. \ref{f1}(f). Detailed calculations and discussions about Q and mode volume are shown in the Supporting Information. Since the designed topological nanocavity has such high Q and small mode volume, it can be used for the investigation of CQED.

\begin{figure}
\centering
\includegraphics[scale=0.43]{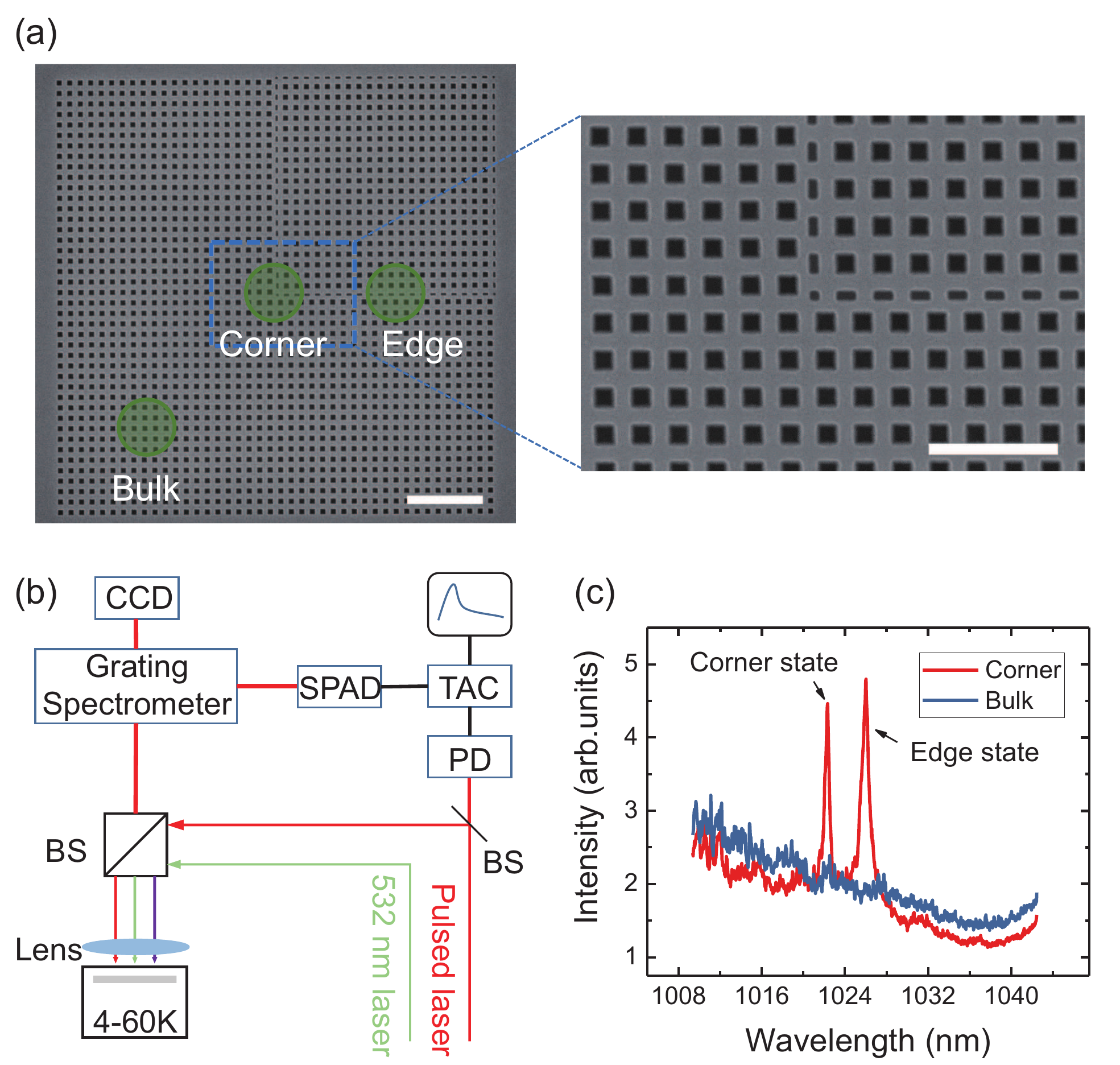}
\caption{ (a) SEM image of a fabricated topological cavity with a scale bar of 2 $\mu m$. The inset shows an enlarged image of the blue area with a scale bar of 1 $\mu m$, indicating the location of the corner state. The different excitation positions are represented by green areas. (b) Schematics of measurement setup for PL measurement and time-resolved PL spectroscopy. (c) PL spectra collected from different positions of the cavity ($a=$ 285 nm, $D=0.61a$ and $g=$ 30 nm) with a pump power of 350 $\mu W$. The red (blue) line is the PL spectrum collected from the corner (bulk PhC). The two peaks are identified as corner state and edge state as the arrows indicated.}
\label{f2_v1}
\end{figure}

\section{Experimental results}

We fabricate the designed topological PhC cavities with different geometric parameters into a 150 nm-thick GaAs slab using electron beam lithography followed by inductively coupled plasma and wet etching process. The GaAs slab is grown by molecular beam epitaxy and contains a single layer of InGaAs QDs at the center. The density of QDs is about 30 $\mu m^{-2}$, which is low enough to investigate the coupling between single QDs and topological cavity. The scanning electron microscope (SEM) image of a fabricated cavity is shown in Fig. \ref{f2_v1}(a). The inset is an enlarged SEM image around the location of the corner state. The fabricated square air holes are not perfect because of fabrication imperfection. However, the corner state can still exist with the slight shape perturbation, as long as there is no topological phase transition happens.

Then we perform the confocal micro-PL measurement at low temperature using a liquid helium flow cryostat as shown in Fig. \ref{f2_v1}(b). The topological cavity is excited by a continuous laser with wavelength of 532 nm. The PL signal is collected by a grating spectrometer and detected with a liquid-nitrogen-cooled charge coupled device camera with a spectral resolution of 60 $\mu eV$. Fig. \ref{f2_v1}(c) shows the collected PL spectra when the cavity with $g=30$ nm is excited at different positions of the PhC slab with high excitation power, which are indicated in Fig. \ref{f2_v1}{a}. Two peaks are observed in the PL spectrum when excited around the corner (red line in Fig. \ref{f2_v1}(c)) while disappear in the bulk (blue line in Fig. \ref{f2_v1}(c)). According to the electric field profiles (Fig. \ref{f1}(c) and (d)), the edge state will also be excited besides the corner state when the laser is focused at the corner. Furthermore, the energy difference between the edge state and corner state will decrease with increasing $g$, enabling the observation of edge state close to the corner state in the spectrum. So the two peaks originate from the corner state and edge state. Thereinto the PL peak with short wavelength is identified as the corner state while the peak with long wavelength is the edge state. The Q of the corner state and the edge state are about 1900 and 1200, respectively. The coexistence of corner and edge state with high Q provides a new platform to integrate the waveguide and cavity on a single chip.

\begin{figure}
\centering
\includegraphics[scale=0.43]{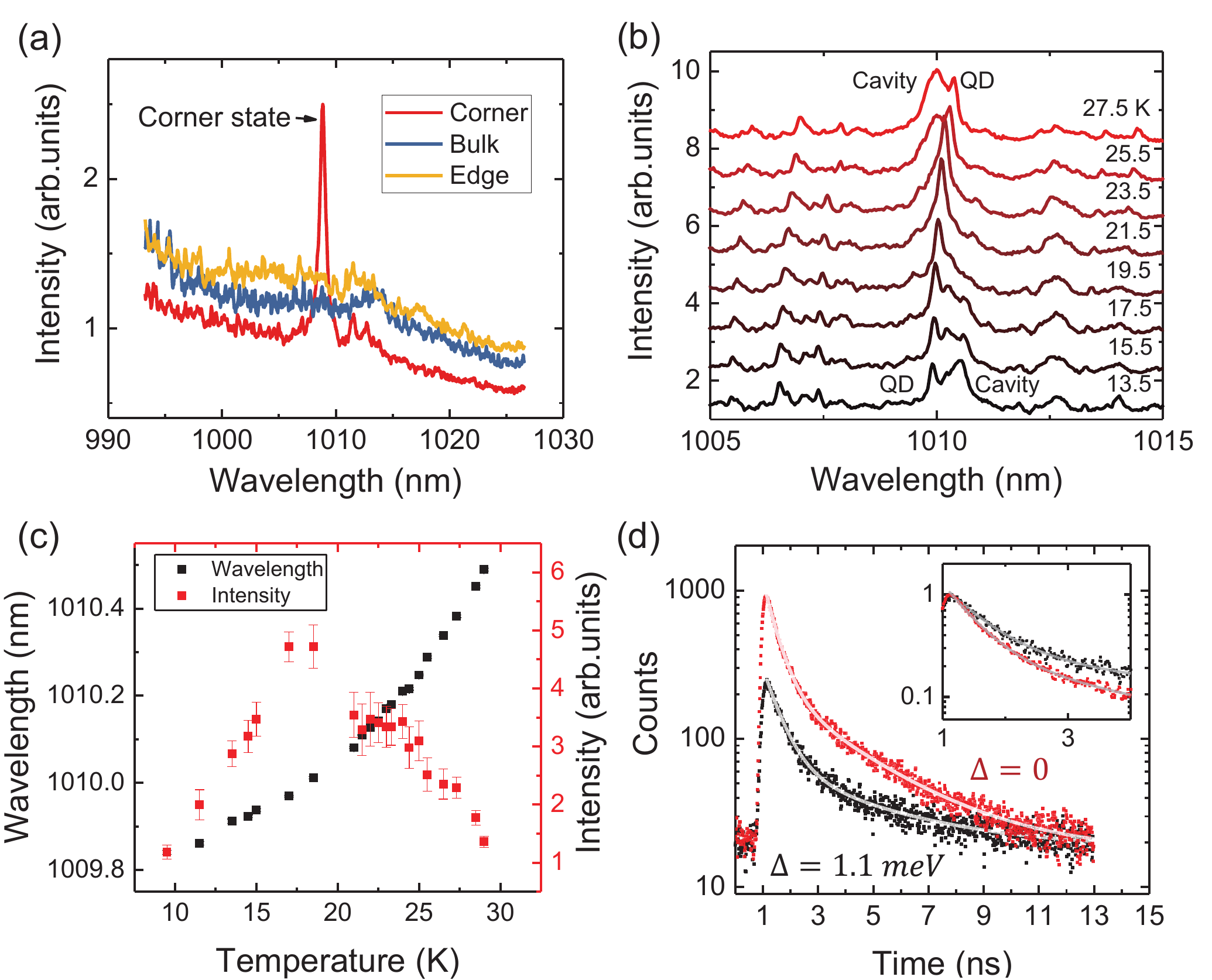}
  \caption{ (a) PL spectra collected when a topological cavity with $a=$ 280 nm and $g=$ 20 nm is excited at different positions. The corner state is identified. (b) PL spectra collected when the QDs are tuned across the corner state by temperature. The spectra are shifted for clarity. (c) Fitted intensity and wavelength of the QD indicated in (b). (d) Decay curves for QD tuned on resonance and off resonance with the corner state. Inset shows the normalized decay curves. The black (red) squares are the experimental results with a detuning $\Delta = 1.1$ $meV$ ($\Delta = 0$ ). The fitted results are represented by the grey lines.     }
\label{f3_v1}
\end{figure}

In order to resolve single QD lines and therefore investigate the interaction between QD and corner state, we pump the sample at low-excitation power about tens of $\mu W$. The stronger pump power compared with other CQED experiments \cite{yoshie2004vacuum,hennessy2007quantum,qian2018two,qian2019enhanced} may result from more defects in the samples. We tune the single QD lines across the corner state by temperature and the corner state with linewidth about 730 $\mu eV$ is identified by the PL spectra (as shown in Fig. \ref{f3_v1}(a)). The absence of peak from the edge state may result from the low Q-factor induced by the fabrication imperfection. Fig. \ref{f3_v1}(b) shows the PL spectra while the QD is tuned across the corner state by temperature. A crossing behavior with an obvious enhancement is observed, suggesting the cavity-QD system is in the Purcell regime. The intensity and wavelength fitted with Lorentz lineshape are shown in Fig. \ref{f3_v1}(c). The intensity of the QD is enhanced by a factor of about 4 on resonance with the corner state.

The Purcell enhancement is further studied by means of time-resolved PL spectroscopy. The schematic of measurement setup is shown in Fig. \ref{f2_v1}(b). The pulsed laser centered at 750 nm with a repetition rate of 82.5 MHz is used. It is split to two beams by a beam splitter (BS). One is used for excitation of the QDs, the other is used as synchronization signal detected by a photodiode (PD). The PL signals are filtered by the grating monochrometer with a resolution less than 100 $\mu eV$ then detected by a single-photon avalanche diode (SPAD). The time differences between photon detection events and synchronization pulses provided by the laser are transformed to electrical signals by a time-to-amplitude converter (TAC), which give the decay curve of the QD in topological cavity.

We measured the decay curves for the QD tuned on resonance (red points) and off resonance (black points) with the corner state, as shown in  Fig. \ref{f3_v1}(d). From the inset, it can be clearly seen that the QD decays faster when on resonance. The curves can be well fitted by a biexponential decay function with fast and slow decay components. When the QD is off resonance (black), the fast and slow decay lifetime of QD are about 0.62 ns and 3.85 ns. On resonance (red), they are 0.48 ns and 3.00 ns, respectively. Both fast and slow decay components are changed, so they could be attributed to the two fine-structure components of neutral exciton \cite{Hohenester2009}. The fast and slow decay rate are both enhanced by a factor of about 1.3, which are due to the Purcell effect. Here, limited by the weak enhancement, we didn't take the lifetime of other QDs in positions without patterning the cavity as a reference. However, the Purcell enhancement can be extracted by comparison of the two decay curves with zero detuning and detuning that is much larger than the linewidth of cavity mode \cite{sapienza2010cavity,Laucht_2009}. In the coupling system between QD and cavity, the off-resonant coupling assisted by acoustic phonons generally exists and strongly depends on temperature and detuning \cite{Hohenester2009}. Unfortunately, it is difficult to estimate its influence in our system since both temperature and detuning are changed during the measurement. Even though the effect of phonons is not taken into consideration, the enhancement of PL intensity and the emission rate sufficiently demonstrate the weak coupling between the QD and the corner state. The weak enhancement may result from spatial misalignment and polarization mismatch between QD and corner state. With better spatial and polarization matching, for example, with QDs embedded in the position with the strongest electric field which is around the smallest square at the corner (See Fig. S3 in Supporting Information), the coupling strength can be improved.

\section{Conclusion}

In conclusion, we have demonstrated a CQED system with second-order topological corner state. We designed and fabricated the topological PhC cavity based on 0D topological corner state. The coexistence of edge state and corner state with Q about $10^3$ is observed. The Purcell enhancement of single QD on resonance with the corner state is demonstrated by means of PL spectra and time-resolved PL spectroscopy. The Q of corner state can be further optimized by changing the position and size of the air holes without changing the topology, therefore making the realization of the strong coupling regime in such a cavity-QD regime possible. Our results provide a new platform to investigate CQED combined with topology, offering an approach to topological devices for quantum information processing. Additionally, the coexistence of edge state and corner state with high Q may enable the integration of topological cavity and waveguide on a single chip, with potential applications in the realization of quantum photonic internet.


\begin{acknowledgments}
X. Xie and W. Zhang contributed equally to this work. This work was supported by the National Natural Science Foundation of China (Grants No. 11934019, No.11721404, No. 51761145104, No. 61675228, and No. 11874419), the National key R$\&$D Program of China (Grant No. 2017YFA0303800 and No. 2018YFA0306101), the Key R$\&$D Program of Guangdong Province (Grant No. 2018B030329001), the Strategic Priority Research Program (Grant No. XDB28000000), the Instrument Developing Project (Grant No. YJKYYQ20180036) and the Interdisciplinary Innovation Team of the Chinese Academy of Sciences.
\end{acknowledgments}

%

%


\begin{thebibliography}{76}%
\makeatletter
\providecommand \@ifxundefined [1]{%
 \@ifx{#1\undefined}
}%
\providecommand \@ifnum [1]{%
 \ifnum #1\expandafter \@firstoftwo
 \else \expandafter \@secondoftwo
 \fi
}%
\providecommand \@ifx [1]{%
 \ifx #1\expandafter \@firstoftwo
 \else \expandafter \@secondoftwo
 \fi
}%
\providecommand \natexlab [1]{#1}%
\providecommand \enquote  [1]{``#1''}%
\providecommand \bibnamefont  [1]{#1}%
\providecommand \bibfnamefont [1]{#1}%
\providecommand \citenamefont [1]{#1}%
\providecommand \href@noop [0]{\@secondoftwo}%
\providecommand \href [0]{\begingroup \@sanitize@url \@href}%
\providecommand \@href[1]{\@@startlink{#1}\@@href}%
\providecommand \@@href[1]{\endgroup#1\@@endlink}%
\providecommand \@sanitize@url [0]{\catcode `\\12\catcode `\$12\catcode
  `\&12\catcode `\#12\catcode `\^12\catcode `\_12\catcode `\%12\relax}%
\providecommand \@@startlink[1]{}%
\providecommand \@@endlink[0]{}%
\providecommand \url  [0]{\begingroup\@sanitize@url \@url }%
\providecommand \@url [1]{\endgroup\@href {#1}{\urlprefix }}%
\providecommand \urlprefix  [0]{URL }%
\providecommand \Eprint [0]{\href }%
\providecommand \doibase [0]{http://dx.doi.org/}%
\providecommand \selectlanguage [0]{\@gobble}%
\providecommand \bibinfo  [0]{\@secondoftwo}%
\providecommand \bibfield  [0]{\@secondoftwo}%
\providecommand \translation [1]{[#1]}%
\providecommand \BibitemOpen [0]{}%
\providecommand \bibitemStop [0]{}%
\providecommand \bibitemNoStop [0]{.\EOS\space}%
\providecommand \EOS [0]{\spacefactor3000\relax}%
\providecommand \BibitemShut  [1]{\csname bibitem#1\endcsname}%
\let\auto@bib@innerbib\@empty
\bibitem [{\citenamefont {Vahala}(2003)}]{vahala2003optical}%
  \BibitemOpen
  \bibfield  {author} {\bibinfo {author} {\bibfnamefont {Kerry~J}\ \bibnamefont
  {Vahala}},\ }\bibfield  {title} {\enquote {\bibinfo {title} {Optical
  microcavities},}\ }\href@noop {} {\bibfield  {journal} {\bibinfo  {journal}
  {Nature}\ }\textbf {\bibinfo {volume} {424}},\ \bibinfo {pages} {839}
  (\bibinfo {year} {2003})}\BibitemShut {NoStop}%
\bibitem [{\citenamefont {Khitrova}\ \emph {et~al.}(2006)\citenamefont
  {Khitrova}, \citenamefont {Gibbs}, \citenamefont {Kira}, \citenamefont
  {Koch},\ and\ \citenamefont {Scherer}}]{khitrova2006vacuum}%
  \BibitemOpen
  \bibfield  {author} {\bibinfo {author} {\bibfnamefont {Galina}\ \bibnamefont
  {Khitrova}}, \bibinfo {author} {\bibfnamefont {HM}~\bibnamefont {Gibbs}},
  \bibinfo {author} {\bibfnamefont {M}~\bibnamefont {Kira}}, \bibinfo {author}
  {\bibfnamefont {Stephan~W}\ \bibnamefont {Koch}}, \ and\ \bibinfo {author}
  {\bibfnamefont {Axel}\ \bibnamefont {Scherer}},\ }\bibfield  {title}
  {\enquote {\bibinfo {title} {Vacuum rabi splitting in semiconductors},}\
  }\href@noop {} {\bibfield  {journal} {\bibinfo  {journal} {Nat. Phys.}\
  }\textbf {\bibinfo {volume} {2}},\ \bibinfo {pages} {81} (\bibinfo {year}
  {2006})}\BibitemShut {NoStop}%
\bibitem [{\citenamefont {Lodahl}\ \emph {et~al.}(2015)\citenamefont {Lodahl},
  \citenamefont {Mahmoodian},\ and\ \citenamefont
  {Stobbe}}]{lodahl2015interfacing}%
  \BibitemOpen
  \bibfield  {author} {\bibinfo {author} {\bibfnamefont {Peter}\ \bibnamefont
  {Lodahl}}, \bibinfo {author} {\bibfnamefont {Sahand}\ \bibnamefont
  {Mahmoodian}}, \ and\ \bibinfo {author} {\bibfnamefont {S{\o}ren}\
  \bibnamefont {Stobbe}},\ }\bibfield  {title} {\enquote {\bibinfo {title}
  {Interfacing single photons and single quantum dots with photonic
  nanostructures},}\ }\href@noop {} {\bibfield  {journal} {\bibinfo  {journal}
  {Rev. Mod. Phys.}\ }\textbf {\bibinfo {volume} {87}},\ \bibinfo {pages} {347}
  (\bibinfo {year} {2015})}\BibitemShut {NoStop}%
\bibitem [{\citenamefont {Imamoglu}\ \emph {et~al.}(1999)\citenamefont
  {Imamoglu}, \citenamefont {Awschalom}, \citenamefont {Burkard}, \citenamefont
  {DiVincenzo}, \citenamefont {Loss}, \citenamefont {Sherwin}, \citenamefont
  {Small} \emph {et~al.}}]{imamog1999quantum}%
  \BibitemOpen
  \bibfield  {author} {\bibinfo {author} {\bibfnamefont {A}~\bibnamefont
  {Imamoglu}}, \bibinfo {author} {\bibfnamefont {David~D}\ \bibnamefont
  {Awschalom}}, \bibinfo {author} {\bibfnamefont {Guido}\ \bibnamefont
  {Burkard}}, \bibinfo {author} {\bibfnamefont {David~P}\ \bibnamefont
  {DiVincenzo}}, \bibinfo {author} {\bibfnamefont {Daniel}\ \bibnamefont
  {Loss}}, \bibinfo {author} {\bibfnamefont {M}~\bibnamefont {Sherwin}},
  \bibinfo {author} {\bibfnamefont {A}~\bibnamefont {Small}},  \emph {et~al.},\
  }\bibfield  {title} {\enquote {\bibinfo {title} {Quantum information
  processing using quantum dot spins and cavity qed},}\ }\href@noop {}
  {\bibfield  {journal} {\bibinfo  {journal} {Phys. Rev. Lett.}\ }\textbf
  {\bibinfo {volume} {83}},\ \bibinfo {pages} {4204} (\bibinfo {year}
  {1999})}\BibitemShut {NoStop}%
\bibitem [{\citenamefont {Purcell}(1995)}]{purcell1995spontaneous}%
  \BibitemOpen
  \bibfield  {author} {\bibinfo {author} {\bibfnamefont {Edward~Mills}\
  \bibnamefont {Purcell}},\ }\bibfield  {title} {\enquote {\bibinfo {title}
  {Spontaneous emission probabilities at radio frequencies},}\ }in\ \href@noop
  {} {\emph {\bibinfo {booktitle} {Confined electrons and photons}}}\ (\bibinfo
   {publisher} {Springer},\ \bibinfo {year} {1995})\ pp.\ \bibinfo {pages}
  {839--839}\BibitemShut {NoStop}%
\bibitem [{\citenamefont {Englund}\ \emph {et~al.}(2005)\citenamefont
  {Englund}, \citenamefont {Fattal}, \citenamefont {Waks}, \citenamefont
  {Solomon}, \citenamefont {Zhang}, \citenamefont {Nakaoka}, \citenamefont
  {Arakawa}, \citenamefont {Yamamoto},\ and\ \citenamefont
  {Vu{\v{c}}kovi{\'c}}}]{englund2005controlling}%
  \BibitemOpen
  \bibfield  {author} {\bibinfo {author} {\bibfnamefont {Dirk}\ \bibnamefont
  {Englund}}, \bibinfo {author} {\bibfnamefont {David}\ \bibnamefont {Fattal}},
  \bibinfo {author} {\bibfnamefont {Edo}\ \bibnamefont {Waks}}, \bibinfo
  {author} {\bibfnamefont {Glenn}\ \bibnamefont {Solomon}}, \bibinfo {author}
  {\bibfnamefont {Bingyang}\ \bibnamefont {Zhang}}, \bibinfo {author}
  {\bibfnamefont {Toshihiro}\ \bibnamefont {Nakaoka}}, \bibinfo {author}
  {\bibfnamefont {Yasuhiko}\ \bibnamefont {Arakawa}}, \bibinfo {author}
  {\bibfnamefont {Yoshihisa}\ \bibnamefont {Yamamoto}}, \ and\ \bibinfo
  {author} {\bibfnamefont {Jelena}\ \bibnamefont {Vu{\v{c}}kovi{\'c}}},\
  }\bibfield  {title} {\enquote {\bibinfo {title} {Controlling the spontaneous
  emission rate of single quantum dots in a two-dimensional photonic
  crystal},}\ }\href@noop {} {\bibfield  {journal} {\bibinfo  {journal} {Phys.
  Rev. Lett.}\ }\textbf {\bibinfo {volume} {95}},\ \bibinfo {pages} {013904}
  (\bibinfo {year} {2005})}\BibitemShut {NoStop}%
\bibitem [{\citenamefont {Laucht}\ \emph {et~al.}(2009)\citenamefont {Laucht},
  \citenamefont {Hofbauer}, \citenamefont {Hauke}, \citenamefont {Angele},
  \citenamefont {Stobbe}, \citenamefont {Kaniber}, \citenamefont {Böhm},
  \citenamefont {Lodahl}, \citenamefont {Amann},\ and\ \citenamefont
  {Finley}}]{Laucht_2009}%
  \BibitemOpen
  \bibfield  {author} {\bibinfo {author} {\bibfnamefont {A}~\bibnamefont
  {Laucht}}, \bibinfo {author} {\bibfnamefont {F}~\bibnamefont {Hofbauer}},
  \bibinfo {author} {\bibfnamefont {N}~\bibnamefont {Hauke}}, \bibinfo {author}
  {\bibfnamefont {J}~\bibnamefont {Angele}}, \bibinfo {author} {\bibfnamefont
  {S}~\bibnamefont {Stobbe}}, \bibinfo {author} {\bibfnamefont {M}~\bibnamefont
  {Kaniber}}, \bibinfo {author} {\bibfnamefont {G}~\bibnamefont {Böhm}},
  \bibinfo {author} {\bibfnamefont {P}~\bibnamefont {Lodahl}}, \bibinfo
  {author} {\bibfnamefont {M-C}\ \bibnamefont {Amann}}, \ and\ \bibinfo
  {author} {\bibfnamefont {J~J}\ \bibnamefont {Finley}},\ }\bibfield  {title}
  {\enquote {\bibinfo {title} {Electrical control of spontaneous emission and
  strong coupling for a single quantum dot},}\ }\href@noop {} {\bibfield
  {journal} {\bibinfo  {journal} {New J. Phys.}\ }\textbf {\bibinfo {volume}
  {11}},\ \bibinfo {pages} {023034} (\bibinfo {year} {2009})}\BibitemShut
  {NoStop}%
\bibitem [{\citenamefont {Liu}\ \emph {et~al.}(2018)\citenamefont {Liu},
  \citenamefont {Brash}, \citenamefont {O’Hara}, \citenamefont {Martins},
  \citenamefont {Phillips}, \citenamefont {Coles}, \citenamefont {Royall},
  \citenamefont {Clarke}, \citenamefont {Bentham}, \citenamefont {Prtljaga}
  \emph {et~al.}}]{liu2018high}%
  \BibitemOpen
  \bibfield  {author} {\bibinfo {author} {\bibfnamefont {Feng}\ \bibnamefont
  {Liu}}, \bibinfo {author} {\bibfnamefont {Alistair~J}\ \bibnamefont {Brash}},
  \bibinfo {author} {\bibfnamefont {John}\ \bibnamefont {O’Hara}}, \bibinfo
  {author} {\bibfnamefont {Luis~MPP}\ \bibnamefont {Martins}}, \bibinfo
  {author} {\bibfnamefont {Catherine~L}\ \bibnamefont {Phillips}}, \bibinfo
  {author} {\bibfnamefont {Rikki~J}\ \bibnamefont {Coles}}, \bibinfo {author}
  {\bibfnamefont {Benjamin}\ \bibnamefont {Royall}}, \bibinfo {author}
  {\bibfnamefont {Edmund}\ \bibnamefont {Clarke}}, \bibinfo {author}
  {\bibfnamefont {Christopher}\ \bibnamefont {Bentham}}, \bibinfo {author}
  {\bibfnamefont {Nikola}\ \bibnamefont {Prtljaga}},  \emph {et~al.},\
  }\bibfield  {title} {\enquote {\bibinfo {title} {High purcell factor
  generation of indistinguishable on-chip single photons},}\ }\href@noop {}
  {\bibfield  {journal} {\bibinfo  {journal} {Nat. Nanotechnol.}\ }\textbf
  {\bibinfo {volume} {13}},\ \bibinfo {pages} {835} (\bibinfo {year}
  {2018})}\BibitemShut {NoStop}%
\bibitem [{\citenamefont {Chang}\ \emph {et~al.}(2006)\citenamefont {Chang},
  \citenamefont {Chen}, \citenamefont {Chang}, \citenamefont {Hsieh},
  \citenamefont {Chyi},\ and\ \citenamefont {Hsu}}]{Chang2019Efficient}%
  \BibitemOpen
  \bibfield  {author} {\bibinfo {author} {\bibfnamefont {WH}~\bibnamefont
  {Chang}}, \bibinfo {author} {\bibfnamefont {WY}~\bibnamefont {Chen}},
  \bibinfo {author} {\bibfnamefont {HS}~\bibnamefont {Chang}}, \bibinfo
  {author} {\bibfnamefont {TP}~\bibnamefont {Hsieh}}, \bibinfo {author}
  {\bibfnamefont {JI}~\bibnamefont {Chyi}}, \ and\ \bibinfo {author}
  {\bibfnamefont {TM}~\bibnamefont {Hsu}},\ }\bibfield  {title} {\enquote
  {\bibinfo {title} {Efficient single-photon sources based on low-density
  quantum dots in photonic-crystal nanocavities},}\ }\href@noop {} {\bibfield
  {journal} {\bibinfo  {journal} {Phys. Rev. Lett.}\ }\textbf {\bibinfo
  {volume} {96}},\ \bibinfo {pages} {117401} (\bibinfo {year}
  {2006})}\BibitemShut {NoStop}%
\bibitem [{\citenamefont {Strauf}\ \emph {et~al.}(2007)\citenamefont {Strauf},
  \citenamefont {Stoltz}, \citenamefont {Rakher}, \citenamefont {Coldren},
  \citenamefont {Petroff},\ and\ \citenamefont {Bouwmeester}}]{strauf2007high}%
  \BibitemOpen
  \bibfield  {author} {\bibinfo {author} {\bibfnamefont {Stefan}\ \bibnamefont
  {Strauf}}, \bibinfo {author} {\bibfnamefont {Nick~G}\ \bibnamefont {Stoltz}},
  \bibinfo {author} {\bibfnamefont {Matthew~T}\ \bibnamefont {Rakher}},
  \bibinfo {author} {\bibfnamefont {Larry~A}\ \bibnamefont {Coldren}}, \bibinfo
  {author} {\bibfnamefont {Pierre~M}\ \bibnamefont {Petroff}}, \ and\ \bibinfo
  {author} {\bibfnamefont {Dirk}\ \bibnamefont {Bouwmeester}},\ }\bibfield
  {title} {\enquote {\bibinfo {title} {High-frequency single-photon source with
  polarization control},}\ }\href@noop {} {\bibfield  {journal} {\bibinfo
  {journal} {Nat. Photonics}\ }\textbf {\bibinfo {volume} {1}},\ \bibinfo
  {pages} {704} (\bibinfo {year} {2007})}\BibitemShut {NoStop}%
\bibitem [{\citenamefont {Carter}\ \emph {et~al.}(2013)\citenamefont {Carter},
  \citenamefont {Sweeney}, \citenamefont {Kim}, \citenamefont {Kim},
  \citenamefont {Solenov}, \citenamefont {Economou}, \citenamefont {Reinecke},
  \citenamefont {Yang}, \citenamefont {Bracker},\ and\ \citenamefont
  {Gammon}}]{carter2013quantum}%
  \BibitemOpen
  \bibfield  {author} {\bibinfo {author} {\bibfnamefont {Samuel~G}\
  \bibnamefont {Carter}}, \bibinfo {author} {\bibfnamefont {Timothy~M}\
  \bibnamefont {Sweeney}}, \bibinfo {author} {\bibfnamefont {Mijin}\
  \bibnamefont {Kim}}, \bibinfo {author} {\bibfnamefont {Chul~Soo}\
  \bibnamefont {Kim}}, \bibinfo {author} {\bibfnamefont {Dmitry}\ \bibnamefont
  {Solenov}}, \bibinfo {author} {\bibfnamefont {Sophia~E}\ \bibnamefont
  {Economou}}, \bibinfo {author} {\bibfnamefont {Thomas~L}\ \bibnamefont
  {Reinecke}}, \bibinfo {author} {\bibfnamefont {Lily}\ \bibnamefont {Yang}},
  \bibinfo {author} {\bibfnamefont {Allan~S}\ \bibnamefont {Bracker}}, \ and\
  \bibinfo {author} {\bibfnamefont {Daniel}\ \bibnamefont {Gammon}},\
  }\bibfield  {title} {\enquote {\bibinfo {title} {Quantum control of a spin
  qubit coupled to a photonic crystal cavity},}\ }\href@noop {} {\bibfield
  {journal} {\bibinfo  {journal} {Nat. Photonics}\ }\textbf {\bibinfo {volume}
  {7}},\ \bibinfo {pages} {329} (\bibinfo {year} {2013})}\BibitemShut {NoStop}%
\bibitem [{\citenamefont {Sweeney}\ \emph {et~al.}(2014)\citenamefont
  {Sweeney}, \citenamefont {Carter}, \citenamefont {Bracker}, \citenamefont
  {Kim}, \citenamefont {Kim}, \citenamefont {Yang}, \citenamefont {Vora},
  \citenamefont {Brereton}, \citenamefont {Cleveland},\ and\ \citenamefont
  {Gammon}}]{sweeney2014cavity}%
  \BibitemOpen
  \bibfield  {author} {\bibinfo {author} {\bibfnamefont {Timothy~M}\
  \bibnamefont {Sweeney}}, \bibinfo {author} {\bibfnamefont {Samuel~G}\
  \bibnamefont {Carter}}, \bibinfo {author} {\bibfnamefont {Allan~S}\
  \bibnamefont {Bracker}}, \bibinfo {author} {\bibfnamefont {Mijin}\
  \bibnamefont {Kim}}, \bibinfo {author} {\bibfnamefont {Chul~Soo}\
  \bibnamefont {Kim}}, \bibinfo {author} {\bibfnamefont {Lily}\ \bibnamefont
  {Yang}}, \bibinfo {author} {\bibfnamefont {Patrick~M}\ \bibnamefont {Vora}},
  \bibinfo {author} {\bibfnamefont {Peter~G}\ \bibnamefont {Brereton}},
  \bibinfo {author} {\bibfnamefont {Erin~R}\ \bibnamefont {Cleveland}}, \ and\
  \bibinfo {author} {\bibfnamefont {Daniel}\ \bibnamefont {Gammon}},\
  }\bibfield  {title} {\enquote {\bibinfo {title} {Cavity-stimulated raman
  emission from a single quantum dot spin},}\ }\href@noop {} {\bibfield
  {journal} {\bibinfo  {journal} {Nat. Photonics}\ }\textbf {\bibinfo {volume}
  {8}},\ \bibinfo {pages} {442} (\bibinfo {year} {2014})}\BibitemShut {NoStop}%
\bibitem [{\citenamefont {Lon{\v{c}}ar}\ \emph {et~al.}(2002)\citenamefont
  {Lon{\v{c}}ar}, \citenamefont {Yoshie}, \citenamefont {Scherer},
  \citenamefont {Gogna},\ and\ \citenamefont {Qiu}}]{lonvcar2002low}%
  \BibitemOpen
  \bibfield  {author} {\bibinfo {author} {\bibfnamefont {Marko}\ \bibnamefont
  {Lon{\v{c}}ar}}, \bibinfo {author} {\bibfnamefont {Tomoyuki}\ \bibnamefont
  {Yoshie}}, \bibinfo {author} {\bibfnamefont {Axel}\ \bibnamefont {Scherer}},
  \bibinfo {author} {\bibfnamefont {Pawan}\ \bibnamefont {Gogna}}, \ and\
  \bibinfo {author} {\bibfnamefont {Yueming}\ \bibnamefont {Qiu}},\ }\bibfield
  {title} {\enquote {\bibinfo {title} {Low-threshold photonic crystal laser},}\
  }\href@noop {} {\bibfield  {journal} {\bibinfo  {journal} {Appl. Phys.
  Lett.}\ }\textbf {\bibinfo {volume} {81}},\ \bibinfo {pages} {2680--2682}
  (\bibinfo {year} {2002})}\BibitemShut {NoStop}%
\bibitem [{\citenamefont {Srinivasan}\ and\ \citenamefont
  {Painter}(2007)}]{srinivasan2007linear}%
  \BibitemOpen
  \bibfield  {author} {\bibinfo {author} {\bibfnamefont {Kartik}\ \bibnamefont
  {Srinivasan}}\ and\ \bibinfo {author} {\bibfnamefont {Oskar}\ \bibnamefont
  {Painter}},\ }\bibfield  {title} {\enquote {\bibinfo {title} {Linear and
  nonlinear optical spectroscopy of a strongly coupled microdisk--quantum dot
  system},}\ }\href@noop {} {\bibfield  {journal} {\bibinfo  {journal}
  {Nature}\ }\textbf {\bibinfo {volume} {450}},\ \bibinfo {pages} {862}
  (\bibinfo {year} {2007})}\BibitemShut {NoStop}%
\bibitem [{\citenamefont {Sapienza}\ \emph {et~al.}(2010)\citenamefont
  {Sapienza}, \citenamefont {Thyrrestrup}, \citenamefont {Stobbe},
  \citenamefont {Garcia}, \citenamefont {Smolka},\ and\ \citenamefont
  {Lodahl}}]{sapienza2010cavity}%
  \BibitemOpen
  \bibfield  {author} {\bibinfo {author} {\bibfnamefont {Luca}\ \bibnamefont
  {Sapienza}}, \bibinfo {author} {\bibfnamefont {Henri}\ \bibnamefont
  {Thyrrestrup}}, \bibinfo {author} {\bibfnamefont {S{\o}ren}\ \bibnamefont
  {Stobbe}}, \bibinfo {author} {\bibfnamefont {Pedro~David}\ \bibnamefont
  {Garcia}}, \bibinfo {author} {\bibfnamefont {Stephan}\ \bibnamefont
  {Smolka}}, \ and\ \bibinfo {author} {\bibfnamefont {Peter}\ \bibnamefont
  {Lodahl}},\ }\bibfield  {title} {\enquote {\bibinfo {title} {Cavity quantum
  electrodynamics with anderson-localized modes},}\ }\href@noop {} {\bibfield
  {journal} {\bibinfo  {journal} {Science}\ }\textbf {\bibinfo {volume}
  {327}},\ \bibinfo {pages} {1352--1355} (\bibinfo {year} {2010})}\BibitemShut
  {NoStop}%
\bibitem [{\citenamefont {Yoshie}\ \emph {et~al.}(2004)\citenamefont {Yoshie},
  \citenamefont {Scherer}, \citenamefont {Hendrickson}, \citenamefont
  {Khitrova}, \citenamefont {Gibbs}, \citenamefont {Rupper}, \citenamefont
  {Ell}, \citenamefont {Shchekin},\ and\ \citenamefont
  {Deppe}}]{yoshie2004vacuum}%
  \BibitemOpen
  \bibfield  {author} {\bibinfo {author} {\bibfnamefont {Tomoyuki}\
  \bibnamefont {Yoshie}}, \bibinfo {author} {\bibfnamefont {Axel}\ \bibnamefont
  {Scherer}}, \bibinfo {author} {\bibfnamefont {J}~\bibnamefont {Hendrickson}},
  \bibinfo {author} {\bibfnamefont {Galina}\ \bibnamefont {Khitrova}}, \bibinfo
  {author} {\bibfnamefont {HM}~\bibnamefont {Gibbs}}, \bibinfo {author}
  {\bibfnamefont {G}~\bibnamefont {Rupper}}, \bibinfo {author} {\bibfnamefont
  {C}~\bibnamefont {Ell}}, \bibinfo {author} {\bibfnamefont {OB}~\bibnamefont
  {Shchekin}}, \ and\ \bibinfo {author} {\bibfnamefont {DG}~\bibnamefont
  {Deppe}},\ }\bibfield  {title} {\enquote {\bibinfo {title} {Vacuum rabi
  splitting with a single quantum dot in a photonic crystal nanocavity},}\
  }\href@noop {} {\bibfield  {journal} {\bibinfo  {journal} {Nature}\ }\textbf
  {\bibinfo {volume} {432}},\ \bibinfo {pages} {200} (\bibinfo {year}
  {2004})}\BibitemShut {NoStop}%
\bibitem [{\citenamefont {Hennessy}\ \emph {et~al.}(2007)\citenamefont
  {Hennessy}, \citenamefont {Badolato}, \citenamefont {Winger}, \citenamefont
  {Gerace}, \citenamefont {Atat{\"u}re}, \citenamefont {Gulde}, \citenamefont
  {F{\"a}lt}, \citenamefont {Hu},\ and\ \citenamefont
  {Imamo{\u{g}}lu}}]{hennessy2007quantum}%
  \BibitemOpen
  \bibfield  {author} {\bibinfo {author} {\bibfnamefont {Kevin}\ \bibnamefont
  {Hennessy}}, \bibinfo {author} {\bibfnamefont {Antonio}\ \bibnamefont
  {Badolato}}, \bibinfo {author} {\bibfnamefont {Martin}\ \bibnamefont
  {Winger}}, \bibinfo {author} {\bibfnamefont {D}~\bibnamefont {Gerace}},
  \bibinfo {author} {\bibfnamefont {Mete}\ \bibnamefont {Atat{\"u}re}},
  \bibinfo {author} {\bibfnamefont {S}~\bibnamefont {Gulde}}, \bibinfo {author}
  {\bibfnamefont {S}~\bibnamefont {F{\"a}lt}}, \bibinfo {author} {\bibfnamefont
  {Evelyn~L}\ \bibnamefont {Hu}}, \ and\ \bibinfo {author} {\bibfnamefont
  {A}~\bibnamefont {Imamo{\u{g}}lu}},\ }\bibfield  {title} {\enquote {\bibinfo
  {title} {Quantum nature of a strongly coupled single quantum dot--cavity
  system},}\ }\href@noop {} {\bibfield  {journal} {\bibinfo  {journal}
  {Nature}\ }\textbf {\bibinfo {volume} {445}},\ \bibinfo {pages} {896}
  (\bibinfo {year} {2007})}\BibitemShut {NoStop}%
\bibitem [{\citenamefont {Brossard}\ \emph {et~al.}(2010)\citenamefont
  {Brossard}, \citenamefont {Xu}, \citenamefont {Williams}, \citenamefont
  {Hadjipanayi}, \citenamefont {Hugues}, \citenamefont {Hopkinson},
  \citenamefont {Wang},\ and\ \citenamefont {Taylor}}]{brossard2010strongly}%
  \BibitemOpen
  \bibfield  {author} {\bibinfo {author} {\bibfnamefont {FSF}\ \bibnamefont
  {Brossard}}, \bibinfo {author} {\bibfnamefont {XL}~\bibnamefont {Xu}},
  \bibinfo {author} {\bibfnamefont {DA}~\bibnamefont {Williams}}, \bibinfo
  {author} {\bibfnamefont {M}~\bibnamefont {Hadjipanayi}}, \bibinfo {author}
  {\bibfnamefont {M}~\bibnamefont {Hugues}}, \bibinfo {author} {\bibfnamefont
  {M}~\bibnamefont {Hopkinson}}, \bibinfo {author} {\bibfnamefont
  {X}~\bibnamefont {Wang}}, \ and\ \bibinfo {author} {\bibfnamefont
  {RA}~\bibnamefont {Taylor}},\ }\bibfield  {title} {\enquote {\bibinfo {title}
  {Strongly coupled single quantum dot in a photonic crystal waveguide
  cavity},}\ }\href@noop {} {\bibfield  {journal} {\bibinfo  {journal} {Appl.
  Phys. Lett.}\ }\textbf {\bibinfo {volume} {97}},\ \bibinfo {pages} {111101}
  (\bibinfo {year} {2010})}\BibitemShut {NoStop}%
\bibitem [{\citenamefont {Qian}\ \emph {et~al.}(2018)\citenamefont {Qian},
  \citenamefont {Wu}, \citenamefont {Song}, \citenamefont {Peng}, \citenamefont
  {Xie}, \citenamefont {Yang}, \citenamefont {Xiao}, \citenamefont {Steer},
  \citenamefont {Thayne}, \citenamefont {Tang} \emph {et~al.}}]{qian2018two}%
  \BibitemOpen
  \bibfield  {author} {\bibinfo {author} {\bibfnamefont {Chenjiang}\
  \bibnamefont {Qian}}, \bibinfo {author} {\bibfnamefont {Shiyao}\ \bibnamefont
  {Wu}}, \bibinfo {author} {\bibfnamefont {Feilong}\ \bibnamefont {Song}},
  \bibinfo {author} {\bibfnamefont {Kai}\ \bibnamefont {Peng}}, \bibinfo
  {author} {\bibfnamefont {Xin}\ \bibnamefont {Xie}}, \bibinfo {author}
  {\bibfnamefont {Jingnan}\ \bibnamefont {Yang}}, \bibinfo {author}
  {\bibfnamefont {Shan}\ \bibnamefont {Xiao}}, \bibinfo {author} {\bibfnamefont
  {Matthew~J}\ \bibnamefont {Steer}}, \bibinfo {author} {\bibfnamefont
  {Iain~G}\ \bibnamefont {Thayne}}, \bibinfo {author} {\bibfnamefont
  {Chengchun}\ \bibnamefont {Tang}},  \emph {et~al.},\ }\bibfield  {title}
  {\enquote {\bibinfo {title} {Two-photon rabi splitting in a coupled system of
  a nanocavity and exciton complexes},}\ }\href@noop {} {\bibfield  {journal}
  {\bibinfo  {journal} {Phys. Rev. Lett.}\ }\textbf {\bibinfo {volume} {120}},\
  \bibinfo {pages} {213901} (\bibinfo {year} {2018})}\BibitemShut {NoStop}%
\bibitem [{\citenamefont {Qian}\ \emph {et~al.}(2019)\citenamefont {Qian},
  \citenamefont {Xie}, \citenamefont {Yang}, \citenamefont {Peng},
  \citenamefont {Wu}, \citenamefont {Song}, \citenamefont {Sun}, \citenamefont
  {Dang}, \citenamefont {Yu}, \citenamefont {Steer} \emph
  {et~al.}}]{qian2019enhanced}%
  \BibitemOpen
  \bibfield  {author} {\bibinfo {author} {\bibfnamefont {Chenjiang}\
  \bibnamefont {Qian}}, \bibinfo {author} {\bibfnamefont {Xin}\ \bibnamefont
  {Xie}}, \bibinfo {author} {\bibfnamefont {Jingnan}\ \bibnamefont {Yang}},
  \bibinfo {author} {\bibfnamefont {Kai}\ \bibnamefont {Peng}}, \bibinfo
  {author} {\bibfnamefont {Shiyao}\ \bibnamefont {Wu}}, \bibinfo {author}
  {\bibfnamefont {Feilong}\ \bibnamefont {Song}}, \bibinfo {author}
  {\bibfnamefont {Sibai}\ \bibnamefont {Sun}}, \bibinfo {author} {\bibfnamefont
  {Jianchen}\ \bibnamefont {Dang}}, \bibinfo {author} {\bibfnamefont {Yang}\
  \bibnamefont {Yu}}, \bibinfo {author} {\bibfnamefont {Matthew~J}\
  \bibnamefont {Steer}},  \emph {et~al.},\ }\bibfield  {title} {\enquote
  {\bibinfo {title} {Enhanced strong interaction between nanocavities and
  p-shell excitons beyond the dipole approximation},}\ }\href@noop {}
  {\bibfield  {journal} {\bibinfo  {journal} {Phys. Rev. Lett.}\ }\textbf
  {\bibinfo {volume} {122}},\ \bibinfo {pages} {087401} (\bibinfo {year}
  {2019})}\BibitemShut {NoStop}%
\bibitem [{\citenamefont {Khanikaev}\ and\ \citenamefont
  {Shvets}(2017)}]{khanikaev2017two}%
  \BibitemOpen
  \bibfield  {author} {\bibinfo {author} {\bibfnamefont {Alexander~B}\
  \bibnamefont {Khanikaev}}\ and\ \bibinfo {author} {\bibfnamefont {Gennady}\
  \bibnamefont {Shvets}},\ }\bibfield  {title} {\enquote {\bibinfo {title}
  {Two-dimensional topological photonics},}\ }\href@noop {} {\bibfield
  {journal} {\bibinfo  {journal} {Nat. Photonics}\ }\textbf {\bibinfo {volume}
  {11}},\ \bibinfo {pages} {763} (\bibinfo {year} {2017})}\BibitemShut
  {NoStop}%
\bibitem [{\citenamefont {Ozawa}\ \emph {et~al.}(2019)\citenamefont {Ozawa},
  \citenamefont {Price}, \citenamefont {Amo}, \citenamefont {Goldman},
  \citenamefont {Hafezi}, \citenamefont {Lu}, \citenamefont {Rechtsman},
  \citenamefont {Schuster}, \citenamefont {Simon}, \citenamefont {Zilberberg}
  \emph {et~al.}}]{ozawa2019topological}%
  \BibitemOpen
  \bibfield  {author} {\bibinfo {author} {\bibfnamefont {Tomoki}\ \bibnamefont
  {Ozawa}}, \bibinfo {author} {\bibfnamefont {Hannah~M}\ \bibnamefont {Price}},
  \bibinfo {author} {\bibfnamefont {Alberto}\ \bibnamefont {Amo}}, \bibinfo
  {author} {\bibfnamefont {Nathan}\ \bibnamefont {Goldman}}, \bibinfo {author}
  {\bibfnamefont {Mohammad}\ \bibnamefont {Hafezi}}, \bibinfo {author}
  {\bibfnamefont {Ling}\ \bibnamefont {Lu}}, \bibinfo {author} {\bibfnamefont
  {Mikael~C}\ \bibnamefont {Rechtsman}}, \bibinfo {author} {\bibfnamefont
  {David}\ \bibnamefont {Schuster}}, \bibinfo {author} {\bibfnamefont
  {Jonathan}\ \bibnamefont {Simon}}, \bibinfo {author} {\bibfnamefont {Oded}\
  \bibnamefont {Zilberberg}},  \emph {et~al.},\ }\bibfield  {title} {\enquote
  {\bibinfo {title} {{Topological photonics}},}\ }\href@noop {} {\bibfield
  {journal} {\bibinfo  {journal} {Rev. Mod. Phys.}\ }\textbf {\bibinfo {volume}
  {91}},\ \bibinfo {pages} {015006} (\bibinfo {year} {2019})}\BibitemShut
  {NoStop}%
\bibitem [{\citenamefont {Lu}\ \emph {et~al.}(2014)\citenamefont {Lu},
  \citenamefont {Joannopoulos},\ and\ \citenamefont
  {Solja{\v{c}}i{\'c}}}]{lu2014topological}%
  \BibitemOpen
  \bibfield  {author} {\bibinfo {author} {\bibfnamefont {Ling}\ \bibnamefont
  {Lu}}, \bibinfo {author} {\bibfnamefont {John~D}\ \bibnamefont
  {Joannopoulos}}, \ and\ \bibinfo {author} {\bibfnamefont {Marin}\
  \bibnamefont {Solja{\v{c}}i{\'c}}},\ }\bibfield  {title} {\enquote {\bibinfo
  {title} {Topological photonics},}\ }\href@noop {} {\bibfield  {journal}
  {\bibinfo  {journal} {Nat. Photonics}\ }\textbf {\bibinfo {volume} {8}},\
  \bibinfo {pages} {821} (\bibinfo {year} {2014})}\BibitemShut {NoStop}%
\bibitem [{\citenamefont {Wu}\ and\ \citenamefont {Hu}(2015)}]{wu2015scheme}%
  \BibitemOpen
  \bibfield  {author} {\bibinfo {author} {\bibfnamefont {Long-Hua}\
  \bibnamefont {Wu}}\ and\ \bibinfo {author} {\bibfnamefont {Xiao}\
  \bibnamefont {Hu}},\ }\bibfield  {title} {\enquote {\bibinfo {title} {Scheme
  for achieving a topological photonic crystal by using dielectric material},}\
  }\href@noop {} {\bibfield  {journal} {\bibinfo  {journal} {Phys. Rev. Lett.}\
  }\textbf {\bibinfo {volume} {114}},\ \bibinfo {pages} {223901} (\bibinfo
  {year} {2015})}\BibitemShut {NoStop}%
\bibitem [{\citenamefont {Blanco-Redondo}\ \emph {et~al.}(2016)\citenamefont
  {Blanco-Redondo}, \citenamefont {Andonegui}, \citenamefont {Collins},
  \citenamefont {Harari}, \citenamefont {Lumer}, \citenamefont {Rechtsman},
  \citenamefont {Eggleton},\ and\ \citenamefont
  {Segev}}]{blanco2016topological}%
  \BibitemOpen
  \bibfield  {author} {\bibinfo {author} {\bibfnamefont {Andrea}\ \bibnamefont
  {Blanco-Redondo}}, \bibinfo {author} {\bibfnamefont {Imanol}\ \bibnamefont
  {Andonegui}}, \bibinfo {author} {\bibfnamefont {Matthew~J}\ \bibnamefont
  {Collins}}, \bibinfo {author} {\bibfnamefont {Gal}\ \bibnamefont {Harari}},
  \bibinfo {author} {\bibfnamefont {Yaakov}\ \bibnamefont {Lumer}}, \bibinfo
  {author} {\bibfnamefont {Mikael~C}\ \bibnamefont {Rechtsman}}, \bibinfo
  {author} {\bibfnamefont {Benjamin~J}\ \bibnamefont {Eggleton}}, \ and\
  \bibinfo {author} {\bibfnamefont {Mordechai}\ \bibnamefont {Segev}},\
  }\bibfield  {title} {\enquote {\bibinfo {title} {Topological optical
  waveguiding in silicon and the transition between topological and trivial
  defect states},}\ }\href@noop {} {\bibfield  {journal} {\bibinfo  {journal}
  {Phys. Rev. Lett.}\ }\textbf {\bibinfo {volume} {116}},\ \bibinfo {pages}
  {163901} (\bibinfo {year} {2016})}\BibitemShut {NoStop}%
\bibitem [{\citenamefont {Rechtsman}\ \emph
  {et~al.}(2013{\natexlab{a}})\citenamefont {Rechtsman}, \citenamefont
  {Zeuner}, \citenamefont {Plotnik}, \citenamefont {Lumer}, \citenamefont
  {Podolsky}, \citenamefont {Dreisow}, \citenamefont {Nolte}, \citenamefont
  {Segev},\ and\ \citenamefont {Szameit}}]{rechtsman2013photonic}%
  \BibitemOpen
  \bibfield  {author} {\bibinfo {author} {\bibfnamefont {Mikael~C}\
  \bibnamefont {Rechtsman}}, \bibinfo {author} {\bibfnamefont {Julia~M}\
  \bibnamefont {Zeuner}}, \bibinfo {author} {\bibfnamefont {Yonatan}\
  \bibnamefont {Plotnik}}, \bibinfo {author} {\bibfnamefont {Yaakov}\
  \bibnamefont {Lumer}}, \bibinfo {author} {\bibfnamefont {Daniel}\
  \bibnamefont {Podolsky}}, \bibinfo {author} {\bibfnamefont {Felix}\
  \bibnamefont {Dreisow}}, \bibinfo {author} {\bibfnamefont {Stefan}\
  \bibnamefont {Nolte}}, \bibinfo {author} {\bibfnamefont {Mordechai}\
  \bibnamefont {Segev}}, \ and\ \bibinfo {author} {\bibfnamefont {Alexander}\
  \bibnamefont {Szameit}},\ }\bibfield  {title} {\enquote {\bibinfo {title}
  {Photonic floquet topological insulators},}\ }\href@noop {} {\bibfield
  {journal} {\bibinfo  {journal} {Nature}\ }\textbf {\bibinfo {volume} {496}},\
  \bibinfo {pages} {196} (\bibinfo {year} {2013}{\natexlab{a}})}\BibitemShut
  {NoStop}%
\bibitem [{\citenamefont {Rechtsman}\ \emph
  {et~al.}(2013{\natexlab{b}})\citenamefont {Rechtsman}, \citenamefont
  {Plotnik}, \citenamefont {Zeuner}, \citenamefont {Song}, \citenamefont
  {Chen}, \citenamefont {Szameit},\ and\ \citenamefont
  {Segev}}]{rechtsman2013topological}%
  \BibitemOpen
  \bibfield  {author} {\bibinfo {author} {\bibfnamefont {Mikael~C}\
  \bibnamefont {Rechtsman}}, \bibinfo {author} {\bibfnamefont {Yonatan}\
  \bibnamefont {Plotnik}}, \bibinfo {author} {\bibfnamefont {Julia~M}\
  \bibnamefont {Zeuner}}, \bibinfo {author} {\bibfnamefont {Daohong}\
  \bibnamefont {Song}}, \bibinfo {author} {\bibfnamefont {Zhigang}\
  \bibnamefont {Chen}}, \bibinfo {author} {\bibfnamefont {Alexander}\
  \bibnamefont {Szameit}}, \ and\ \bibinfo {author} {\bibfnamefont {Mordechai}\
  \bibnamefont {Segev}},\ }\bibfield  {title} {\enquote {\bibinfo {title}
  {Topological creation and destruction of edge states in photonic graphene},}\
  }\href@noop {} {\bibfield  {journal} {\bibinfo  {journal} {Phys. Rev. Lett.}\
  }\textbf {\bibinfo {volume} {111}},\ \bibinfo {pages} {103901} (\bibinfo
  {year} {2013}{\natexlab{b}})}\BibitemShut {NoStop}%
\bibitem [{\citenamefont {Haldane}\ and\ \citenamefont
  {Raghu}(2008)}]{haldane2008possible}%
  \BibitemOpen
  \bibfield  {author} {\bibinfo {author} {\bibfnamefont {FDM}\ \bibnamefont
  {Haldane}}\ and\ \bibinfo {author} {\bibfnamefont {S}~\bibnamefont {Raghu}},\
  }\bibfield  {title} {\enquote {\bibinfo {title} {Possible realization of
  directional optical waveguides in photonic crystals with broken time-reversal
  symmetry},}\ }\href@noop {} {\bibfield  {journal} {\bibinfo  {journal} {Phys.
  Rev. Lett.}\ }\textbf {\bibinfo {volume} {100}},\ \bibinfo {pages} {013904}
  (\bibinfo {year} {2008})}\BibitemShut {NoStop}%
\bibitem [{\citenamefont {Wang}\ \emph {et~al.}(2009)\citenamefont {Wang},
  \citenamefont {Chong}, \citenamefont {Joannopoulos},\ and\ \citenamefont
  {Solja{\v{c}}i{\'c}}}]{wang2009observation}%
  \BibitemOpen
  \bibfield  {author} {\bibinfo {author} {\bibfnamefont {Zheng}\ \bibnamefont
  {Wang}}, \bibinfo {author} {\bibfnamefont {Yidong}\ \bibnamefont {Chong}},
  \bibinfo {author} {\bibfnamefont {John~D}\ \bibnamefont {Joannopoulos}}, \
  and\ \bibinfo {author} {\bibfnamefont {Marin}\ \bibnamefont
  {Solja{\v{c}}i{\'c}}},\ }\bibfield  {title} {\enquote {\bibinfo {title}
  {Observation of unidirectional backscattering-immune topological
  electromagnetic states},}\ }\href@noop {} {\bibfield  {journal} {\bibinfo
  {journal} {Nature}\ }\textbf {\bibinfo {volume} {461}},\ \bibinfo {pages}
  {772} (\bibinfo {year} {2009})}\BibitemShut {NoStop}%
\bibitem [{\citenamefont {Wang}\ \emph {et~al.}(2008)\citenamefont {Wang},
  \citenamefont {Chong}, \citenamefont {Joannopoulos},\ and\ \citenamefont
  {Solja{\v{c}}i{\'c}}}]{wang2008reflection}%
  \BibitemOpen
  \bibfield  {author} {\bibinfo {author} {\bibfnamefont {Zheng}\ \bibnamefont
  {Wang}}, \bibinfo {author} {\bibfnamefont {YD}~\bibnamefont {Chong}},
  \bibinfo {author} {\bibfnamefont {John~D}\ \bibnamefont {Joannopoulos}}, \
  and\ \bibinfo {author} {\bibfnamefont {Marin}\ \bibnamefont
  {Solja{\v{c}}i{\'c}}},\ }\bibfield  {title} {\enquote {\bibinfo {title}
  {Reflection-free one-way edge modes in a gyromagnetic photonic crystal},}\
  }\href@noop {} {\bibfield  {journal} {\bibinfo  {journal} {Phys. Rev. Lett.}\
  }\textbf {\bibinfo {volume} {100}},\ \bibinfo {pages} {013905} (\bibinfo
  {year} {2008})}\BibitemShut {NoStop}%
\bibitem [{\citenamefont {Fang}\ \emph {et~al.}(2012)\citenamefont {Fang},
  \citenamefont {Yu},\ and\ \citenamefont {Fan}}]{fang2012realizing}%
  \BibitemOpen
  \bibfield  {author} {\bibinfo {author} {\bibfnamefont {Kejie}\ \bibnamefont
  {Fang}}, \bibinfo {author} {\bibfnamefont {Zongfu}\ \bibnamefont {Yu}}, \
  and\ \bibinfo {author} {\bibfnamefont {Shanhui}\ \bibnamefont {Fan}},\
  }\bibfield  {title} {\enquote {\bibinfo {title} {Realizing effective magnetic
  field for photons by controlling the phase of dynamic modulation},}\
  }\href@noop {} {\bibfield  {journal} {\bibinfo  {journal} {Nat. Photonics}\
  }\textbf {\bibinfo {volume} {6}},\ \bibinfo {pages} {782} (\bibinfo {year}
  {2012})}\BibitemShut {NoStop}%
\bibitem [{\citenamefont {Hafezi}\ \emph {et~al.}(2011)\citenamefont {Hafezi},
  \citenamefont {Demler}, \citenamefont {Lukin},\ and\ \citenamefont
  {Taylor}}]{hafezi2011robust}%
  \BibitemOpen
  \bibfield  {author} {\bibinfo {author} {\bibfnamefont {Mohammad}\
  \bibnamefont {Hafezi}}, \bibinfo {author} {\bibfnamefont {Eugene~A}\
  \bibnamefont {Demler}}, \bibinfo {author} {\bibfnamefont {Mikhail~D}\
  \bibnamefont {Lukin}}, \ and\ \bibinfo {author} {\bibfnamefont {Jacob~M}\
  \bibnamefont {Taylor}},\ }\bibfield  {title} {\enquote {\bibinfo {title}
  {Robust optical delay lines with topological protection},}\ }\href@noop {}
  {\bibfield  {journal} {\bibinfo  {journal} {Nat. Phys.}\ }\textbf {\bibinfo
  {volume} {7}},\ \bibinfo {pages} {907} (\bibinfo {year} {2011})}\BibitemShut
  {NoStop}%
\bibitem [{\citenamefont {Khanikaev}\ \emph {et~al.}(2013)\citenamefont
  {Khanikaev}, \citenamefont {Mousavi}, \citenamefont {Tse}, \citenamefont
  {Kargarian}, \citenamefont {MacDonald},\ and\ \citenamefont
  {Shvets}}]{khanikaev2013photonic}%
  \BibitemOpen
  \bibfield  {author} {\bibinfo {author} {\bibfnamefont {Alexander~B}\
  \bibnamefont {Khanikaev}}, \bibinfo {author} {\bibfnamefont {S~Hossein}\
  \bibnamefont {Mousavi}}, \bibinfo {author} {\bibfnamefont {Wang-Kong}\
  \bibnamefont {Tse}}, \bibinfo {author} {\bibfnamefont {Mehdi}\ \bibnamefont
  {Kargarian}}, \bibinfo {author} {\bibfnamefont {Allan~H}\ \bibnamefont
  {MacDonald}}, \ and\ \bibinfo {author} {\bibfnamefont {Gennady}\ \bibnamefont
  {Shvets}},\ }\bibfield  {title} {\enquote {\bibinfo {title} {Photonic
  topological insulators},}\ }\href@noop {} {\bibfield  {journal} {\bibinfo
  {journal} {Nat. Mater.}\ }\textbf {\bibinfo {volume} {12}},\ \bibinfo {pages}
  {233} (\bibinfo {year} {2013})}\BibitemShut {NoStop}%
\bibitem [{\citenamefont {Hafezi}\ \emph {et~al.}(2013)\citenamefont {Hafezi},
  \citenamefont {Mittal}, \citenamefont {Fan}, \citenamefont {Migdall},\ and\
  \citenamefont {Taylor}}]{hafezi2013imaging}%
  \BibitemOpen
  \bibfield  {author} {\bibinfo {author} {\bibfnamefont {Mohammad}\
  \bibnamefont {Hafezi}}, \bibinfo {author} {\bibfnamefont {S}~\bibnamefont
  {Mittal}}, \bibinfo {author} {\bibfnamefont {J}~\bibnamefont {Fan}}, \bibinfo
  {author} {\bibfnamefont {A}~\bibnamefont {Migdall}}, \ and\ \bibinfo {author}
  {\bibfnamefont {JM}~\bibnamefont {Taylor}},\ }\bibfield  {title} {\enquote
  {\bibinfo {title} {Imaging topological edge states in silicon photonics},}\
  }\href@noop {} {\bibfield  {journal} {\bibinfo  {journal} {Nat. Photonics}\
  }\textbf {\bibinfo {volume} {7}},\ \bibinfo {pages} {1001} (\bibinfo {year}
  {2013})}\BibitemShut {NoStop}%
\bibitem [{\citenamefont {Harari}\ \emph {et~al.}(2018)\citenamefont {Harari},
  \citenamefont {Bandres}, \citenamefont {Lumer}, \citenamefont {Rechtsman},
  \citenamefont {Chong}, \citenamefont {Khajavikhan}, \citenamefont
  {Christodoulides},\ and\ \citenamefont {Segev}}]{harari2018topological}%
  \BibitemOpen
  \bibfield  {author} {\bibinfo {author} {\bibfnamefont {Gal}\ \bibnamefont
  {Harari}}, \bibinfo {author} {\bibfnamefont {Miguel~A}\ \bibnamefont
  {Bandres}}, \bibinfo {author} {\bibfnamefont {Yaakov}\ \bibnamefont {Lumer}},
  \bibinfo {author} {\bibfnamefont {Mikael~C}\ \bibnamefont {Rechtsman}},
  \bibinfo {author} {\bibfnamefont {Yi~Dong}\ \bibnamefont {Chong}}, \bibinfo
  {author} {\bibfnamefont {Mercedeh}\ \bibnamefont {Khajavikhan}}, \bibinfo
  {author} {\bibfnamefont {Demetrios~N}\ \bibnamefont {Christodoulides}}, \
  and\ \bibinfo {author} {\bibfnamefont {Mordechai}\ \bibnamefont {Segev}},\
  }\bibfield  {title} {\enquote {\bibinfo {title} {Topological insulator laser:
  theory},}\ }\href@noop {} {\bibfield  {journal} {\bibinfo  {journal}
  {Science}\ }\textbf {\bibinfo {volume} {359}},\ \bibinfo {pages} {eaar4003}
  (\bibinfo {year} {2018})}\BibitemShut {NoStop}%
\bibitem [{\citenamefont {Bandres}\ \emph {et~al.}(2018)\citenamefont
  {Bandres}, \citenamefont {Wittek}, \citenamefont {Harari}, \citenamefont
  {Parto}, \citenamefont {Ren}, \citenamefont {Segev}, \citenamefont
  {Christodoulides},\ and\ \citenamefont
  {Khajavikhan}}]{bandres2018topological}%
  \BibitemOpen
  \bibfield  {author} {\bibinfo {author} {\bibfnamefont {Miguel~A}\
  \bibnamefont {Bandres}}, \bibinfo {author} {\bibfnamefont {Steffen}\
  \bibnamefont {Wittek}}, \bibinfo {author} {\bibfnamefont {Gal}\ \bibnamefont
  {Harari}}, \bibinfo {author} {\bibfnamefont {Midya}\ \bibnamefont {Parto}},
  \bibinfo {author} {\bibfnamefont {Jinhan}\ \bibnamefont {Ren}}, \bibinfo
  {author} {\bibfnamefont {Mordechai}\ \bibnamefont {Segev}}, \bibinfo {author}
  {\bibfnamefont {Demetrios~N}\ \bibnamefont {Christodoulides}}, \ and\
  \bibinfo {author} {\bibfnamefont {Mercedeh}\ \bibnamefont {Khajavikhan}},\
  }\bibfield  {title} {\enquote {\bibinfo {title} {Topological insulator laser:
  Experiments},}\ }\href@noop {} {\bibfield  {journal} {\bibinfo  {journal}
  {Science}\ }\textbf {\bibinfo {volume} {359}},\ \bibinfo {pages} {eaar4005}
  (\bibinfo {year} {2018})}\BibitemShut {NoStop}%
\bibitem [{\citenamefont {Zhao}\ \emph {et~al.}(2018)\citenamefont {Zhao},
  \citenamefont {Miao}, \citenamefont {Teimourpour}, \citenamefont {Malzard},
  \citenamefont {El-Ganainy}, \citenamefont {Schomerus},\ and\ \citenamefont
  {Feng}}]{zhao2018topological}%
  \BibitemOpen
  \bibfield  {author} {\bibinfo {author} {\bibfnamefont {Han}\ \bibnamefont
  {Zhao}}, \bibinfo {author} {\bibfnamefont {Pei}\ \bibnamefont {Miao}},
  \bibinfo {author} {\bibfnamefont {Mohammad~H}\ \bibnamefont {Teimourpour}},
  \bibinfo {author} {\bibfnamefont {Simon}\ \bibnamefont {Malzard}}, \bibinfo
  {author} {\bibfnamefont {Ramy}\ \bibnamefont {El-Ganainy}}, \bibinfo {author}
  {\bibfnamefont {Henning}\ \bibnamefont {Schomerus}}, \ and\ \bibinfo {author}
  {\bibfnamefont {Liang}\ \bibnamefont {Feng}},\ }\bibfield  {title} {\enquote
  {\bibinfo {title} {Topological hybrid silicon microlasers},}\ }\href@noop {}
  {\bibfield  {journal} {\bibinfo  {journal} {Nat. Commun.}\ }\textbf {\bibinfo
  {volume} {9}},\ \bibinfo {pages} {981} (\bibinfo {year} {2018})}\BibitemShut
  {NoStop}%
\bibitem [{\citenamefont {Parto}\ \emph {et~al.}(2018)\citenamefont {Parto},
  \citenamefont {Wittek}, \citenamefont {Hodaei}, \citenamefont {Harari},
  \citenamefont {Bandres}, \citenamefont {Ren}, \citenamefont {Rechtsman},
  \citenamefont {Segev}, \citenamefont {Christodoulides},\ and\ \citenamefont
  {Khajavikhan}}]{parto2018edge}%
  \BibitemOpen
  \bibfield  {author} {\bibinfo {author} {\bibfnamefont {Midya}\ \bibnamefont
  {Parto}}, \bibinfo {author} {\bibfnamefont {Steffen}\ \bibnamefont {Wittek}},
  \bibinfo {author} {\bibfnamefont {Hossein}\ \bibnamefont {Hodaei}}, \bibinfo
  {author} {\bibfnamefont {Gal}\ \bibnamefont {Harari}}, \bibinfo {author}
  {\bibfnamefont {Miguel~A}\ \bibnamefont {Bandres}}, \bibinfo {author}
  {\bibfnamefont {Jinhan}\ \bibnamefont {Ren}}, \bibinfo {author}
  {\bibfnamefont {Mikael~C}\ \bibnamefont {Rechtsman}}, \bibinfo {author}
  {\bibfnamefont {Mordechai}\ \bibnamefont {Segev}}, \bibinfo {author}
  {\bibfnamefont {Demetrios~N}\ \bibnamefont {Christodoulides}}, \ and\
  \bibinfo {author} {\bibfnamefont {Mercedeh}\ \bibnamefont {Khajavikhan}},\
  }\bibfield  {title} {\enquote {\bibinfo {title} {Edge-mode lasing in 1d
  topological active arrays},}\ }\href@noop {} {\bibfield  {journal} {\bibinfo
  {journal} {Phys. Rev. Lett.}\ }\textbf {\bibinfo {volume} {120}},\ \bibinfo
  {pages} {113901} (\bibinfo {year} {2018})}\BibitemShut {NoStop}%
\bibitem [{\citenamefont {Ota}\ \emph {et~al.}(2018)\citenamefont {Ota},
  \citenamefont {Katsumi}, \citenamefont {Watanabe}, \citenamefont {Iwamoto},\
  and\ \citenamefont {Arakawa}}]{ota2018topological}%
  \BibitemOpen
  \bibfield  {author} {\bibinfo {author} {\bibfnamefont {Yasutomo}\
  \bibnamefont {Ota}}, \bibinfo {author} {\bibfnamefont {Ryota}\ \bibnamefont
  {Katsumi}}, \bibinfo {author} {\bibfnamefont {Katsuyuki}\ \bibnamefont
  {Watanabe}}, \bibinfo {author} {\bibfnamefont {Satoshi}\ \bibnamefont
  {Iwamoto}}, \ and\ \bibinfo {author} {\bibfnamefont {Yasuhiko}\ \bibnamefont
  {Arakawa}},\ }\bibfield  {title} {\enquote {\bibinfo {title} {Topological
  photonic crystal nanocavity laser},}\ }\href@noop {} {\bibfield  {journal}
  {\bibinfo  {journal} {Commun. Phys.}\ }\textbf {\bibinfo {volume} {1}},\
  \bibinfo {pages} {86} (\bibinfo {year} {2018})}\BibitemShut {NoStop}%
\bibitem [{\citenamefont {Bahari}\ \emph {et~al.}(2017)\citenamefont {Bahari},
  \citenamefont {Ndao}, \citenamefont {Vallini}, \citenamefont {El~Amili},
  \citenamefont {Fainman},\ and\ \citenamefont
  {Kant{\'e}}}]{bahari2017nonreciprocal}%
  \BibitemOpen
  \bibfield  {author} {\bibinfo {author} {\bibfnamefont {Babak}\ \bibnamefont
  {Bahari}}, \bibinfo {author} {\bibfnamefont {Abdoulaye}\ \bibnamefont
  {Ndao}}, \bibinfo {author} {\bibfnamefont {Felipe}\ \bibnamefont {Vallini}},
  \bibinfo {author} {\bibfnamefont {Abdelkrim}\ \bibnamefont {El~Amili}},
  \bibinfo {author} {\bibfnamefont {Yeshaiahu}\ \bibnamefont {Fainman}}, \ and\
  \bibinfo {author} {\bibfnamefont {Boubacar}\ \bibnamefont {Kant{\'e}}},\
  }\bibfield  {title} {\enquote {\bibinfo {title} {Nonreciprocal lasing in
  topological cavities of arbitrary geometries},}\ }\href@noop {} {\bibfield
  {journal} {\bibinfo  {journal} {Science}\ }\textbf {\bibinfo {volume}
  {358}},\ \bibinfo {pages} {636--640} (\bibinfo {year} {2017})}\BibitemShut
  {NoStop}%
\bibitem [{\citenamefont {St-Jean}\ \emph {et~al.}(2017)\citenamefont
  {St-Jean}, \citenamefont {Goblot}, \citenamefont {Galopin}, \citenamefont
  {Lema{\^\i}tre}, \citenamefont {Ozawa}, \citenamefont {Le~Gratiet},
  \citenamefont {Sagnes}, \citenamefont {Bloch},\ and\ \citenamefont
  {Amo}}]{st2017lasing}%
  \BibitemOpen
  \bibfield  {author} {\bibinfo {author} {\bibfnamefont {P}~\bibnamefont
  {St-Jean}}, \bibinfo {author} {\bibfnamefont {V}~\bibnamefont {Goblot}},
  \bibinfo {author} {\bibfnamefont {E}~\bibnamefont {Galopin}}, \bibinfo
  {author} {\bibfnamefont {A}~\bibnamefont {Lema{\^\i}tre}}, \bibinfo {author}
  {\bibfnamefont {T}~\bibnamefont {Ozawa}}, \bibinfo {author} {\bibfnamefont
  {L}~\bibnamefont {Le~Gratiet}}, \bibinfo {author} {\bibfnamefont
  {I}~\bibnamefont {Sagnes}}, \bibinfo {author} {\bibfnamefont {J}~\bibnamefont
  {Bloch}}, \ and\ \bibinfo {author} {\bibfnamefont {A}~\bibnamefont {Amo}},\
  }\bibfield  {title} {\enquote {\bibinfo {title} {Lasing in topological edge
  states of a one-dimensional lattice},}\ }\href@noop {} {\bibfield  {journal}
  {\bibinfo  {journal} {Nat. Photonics}\ }\textbf {\bibinfo {volume} {11}},\
  \bibinfo {pages} {651} (\bibinfo {year} {2017})}\BibitemShut {NoStop}%
\bibitem [{\citenamefont {Ota}\ \emph {et~al.}(2019{\natexlab{a}})\citenamefont
  {Ota}, \citenamefont {Takaka}, \citenamefont {Ozawa}, \citenamefont {Amo},
  \citenamefont {Jia}, \citenamefont {Kante}, \citenamefont {Notomi},
  \citenamefont {Arakawa},\ and\ \citenamefont {Iwamoto}}]{ota2019active}%
  \BibitemOpen
  \bibfield  {author} {\bibinfo {author} {\bibfnamefont {Yasutomo}\
  \bibnamefont {Ota}}, \bibinfo {author} {\bibfnamefont {Kenta}\ \bibnamefont
  {Takaka}}, \bibinfo {author} {\bibfnamefont {Tomoki}\ \bibnamefont {Ozawa}},
  \bibinfo {author} {\bibfnamefont {Alberto}\ \bibnamefont {Amo}}, \bibinfo
  {author} {\bibfnamefont {Zhetao}\ \bibnamefont {Jia}}, \bibinfo {author}
  {\bibfnamefont {Boubacar}\ \bibnamefont {Kante}}, \bibinfo {author}
  {\bibfnamefont {Masaya}\ \bibnamefont {Notomi}}, \bibinfo {author}
  {\bibfnamefont {Yasuhiko}\ \bibnamefont {Arakawa}}, \ and\ \bibinfo {author}
  {\bibfnamefont {Satoshi}\ \bibnamefont {Iwamoto}},\ }\bibfield  {title}
  {\enquote {\bibinfo {title} {Active topological photonics},}\ }\href@noop {}
  {\bibfield  {journal} {\bibinfo  {journal} {arXiv preprint arXiv:1912.05126}\
  } (\bibinfo {year} {2019}{\natexlab{a}})}\BibitemShut {NoStop}%
\bibitem [{\citenamefont {Chen}\ \emph {et~al.}(2014)\citenamefont {Chen},
  \citenamefont {Jiang}, \citenamefont {Chen}, \citenamefont {Zhu},
  \citenamefont {Zhou}, \citenamefont {Dong},\ and\ \citenamefont
  {Chan}}]{chen2014experimental}%
  \BibitemOpen
  \bibfield  {author} {\bibinfo {author} {\bibfnamefont {Wen-Jie}\ \bibnamefont
  {Chen}}, \bibinfo {author} {\bibfnamefont {Shao-Ji}\ \bibnamefont {Jiang}},
  \bibinfo {author} {\bibfnamefont {Xiao-Dong}\ \bibnamefont {Chen}}, \bibinfo
  {author} {\bibfnamefont {Baocheng}\ \bibnamefont {Zhu}}, \bibinfo {author}
  {\bibfnamefont {Lei}\ \bibnamefont {Zhou}}, \bibinfo {author} {\bibfnamefont
  {Jian-Wen}\ \bibnamefont {Dong}}, \ and\ \bibinfo {author} {\bibfnamefont
  {Che~Ting}\ \bibnamefont {Chan}},\ }\bibfield  {title} {\enquote {\bibinfo
  {title} {Experimental realization of photonic topological insulator in a
  uniaxial metacrystal waveguide},}\ }\href@noop {} {\bibfield  {journal}
  {\bibinfo  {journal} {Nat. Commun.}\ }\textbf {\bibinfo {volume} {5}},\
  \bibinfo {pages} {5782} (\bibinfo {year} {2014})}\BibitemShut {NoStop}%
\bibitem [{\citenamefont {Smirnova}\ \emph {et~al.}(2019)\citenamefont
  {Smirnova}, \citenamefont {Leykam}, \citenamefont {Chong},\ and\
  \citenamefont {Kivshar}}]{smirnova2019nonlinear}%
  \BibitemOpen
  \bibfield  {author} {\bibinfo {author} {\bibfnamefont {Daria}\ \bibnamefont
  {Smirnova}}, \bibinfo {author} {\bibfnamefont {Daniel}\ \bibnamefont
  {Leykam}}, \bibinfo {author} {\bibfnamefont {Yidong}\ \bibnamefont {Chong}},
  \ and\ \bibinfo {author} {\bibfnamefont {Yuri}\ \bibnamefont {Kivshar}},\
  }\bibfield  {title} {\enquote {\bibinfo {title} {Nonlinear topological
  photonics},}\ }\href@noop {} {\bibfield  {journal} {\bibinfo  {journal}
  {arXiv preprint arXiv:1912.01784}\ } (\bibinfo {year} {2019})}\BibitemShut
  {NoStop}%
\bibitem [{\citenamefont {Blanco-Redondo}\ \emph {et~al.}(2018)\citenamefont
  {Blanco-Redondo}, \citenamefont {Bell}, \citenamefont {Oren}, \citenamefont
  {Eggleton},\ and\ \citenamefont {Segev}}]{blanco2018topological}%
  \BibitemOpen
  \bibfield  {author} {\bibinfo {author} {\bibfnamefont {Andrea}\ \bibnamefont
  {Blanco-Redondo}}, \bibinfo {author} {\bibfnamefont {Bryn}\ \bibnamefont
  {Bell}}, \bibinfo {author} {\bibfnamefont {Dikla}\ \bibnamefont {Oren}},
  \bibinfo {author} {\bibfnamefont {Benjamin~J}\ \bibnamefont {Eggleton}}, \
  and\ \bibinfo {author} {\bibfnamefont {Mordechai}\ \bibnamefont {Segev}},\
  }\bibfield  {title} {\enquote {\bibinfo {title} {Topological protection of
  biphoton states},}\ }\href@noop {} {\bibfield  {journal} {\bibinfo  {journal}
  {Science}\ }\textbf {\bibinfo {volume} {362}},\ \bibinfo {pages} {568--571}
  (\bibinfo {year} {2018})}\BibitemShut {NoStop}%
\bibitem [{\citenamefont {Tambasco}\ \emph {et~al.}(2018)\citenamefont
  {Tambasco}, \citenamefont {Corrielli}, \citenamefont {Chapman}, \citenamefont
  {Crespi}, \citenamefont {Zilberberg}, \citenamefont {Osellame},\ and\
  \citenamefont {Peruzzo}}]{tambasco2018quantum}%
  \BibitemOpen
  \bibfield  {author} {\bibinfo {author} {\bibfnamefont {Jean-Luc}\
  \bibnamefont {Tambasco}}, \bibinfo {author} {\bibfnamefont {Giacomo}\
  \bibnamefont {Corrielli}}, \bibinfo {author} {\bibfnamefont {Robert~J}\
  \bibnamefont {Chapman}}, \bibinfo {author} {\bibfnamefont {Andrea}\
  \bibnamefont {Crespi}}, \bibinfo {author} {\bibfnamefont {Oded}\ \bibnamefont
  {Zilberberg}}, \bibinfo {author} {\bibfnamefont {Roberto}\ \bibnamefont
  {Osellame}}, \ and\ \bibinfo {author} {\bibfnamefont {Alberto}\ \bibnamefont
  {Peruzzo}},\ }\bibfield  {title} {\enquote {\bibinfo {title} {Quantum
  interference of topological states of light},}\ }\href@noop {} {\bibfield
  {journal} {\bibinfo  {journal} {Sci.Adv.}\ }\textbf {\bibinfo {volume} {4}},\
  \bibinfo {pages} {eaat3187} (\bibinfo {year} {2018})}\BibitemShut {NoStop}%
\bibitem [{\citenamefont {Mittal}\ \emph {et~al.}(2018)\citenamefont {Mittal},
  \citenamefont {Goldschmidt},\ and\ \citenamefont
  {Hafezi}}]{mittal2018topological}%
  \BibitemOpen
  \bibfield  {author} {\bibinfo {author} {\bibfnamefont {Sunil}\ \bibnamefont
  {Mittal}}, \bibinfo {author} {\bibfnamefont {Elizabeth~A}\ \bibnamefont
  {Goldschmidt}}, \ and\ \bibinfo {author} {\bibfnamefont {Mohammad}\
  \bibnamefont {Hafezi}},\ }\bibfield  {title} {\enquote {\bibinfo {title} {A
  topological source of quantum light},}\ }\href@noop {} {\bibfield  {journal}
  {\bibinfo  {journal} {Nature}\ }\textbf {\bibinfo {volume} {561}},\ \bibinfo
  {pages} {502} (\bibinfo {year} {2018})}\BibitemShut {NoStop}%
\bibitem [{\citenamefont {Wang}\ \emph
  {et~al.}(2019{\natexlab{a}})\citenamefont {Wang}, \citenamefont {Pang},
  \citenamefont {Lu}, \citenamefont {Gao}, \citenamefont {Chang}, \citenamefont
  {Qiao}, \citenamefont {Jiao}, \citenamefont {Tang},\ and\ \citenamefont
  {Jin}}]{wang2019topological}%
  \BibitemOpen
  \bibfield  {author} {\bibinfo {author} {\bibfnamefont {Yao}\ \bibnamefont
  {Wang}}, \bibinfo {author} {\bibfnamefont {Xiao-Ling}\ \bibnamefont {Pang}},
  \bibinfo {author} {\bibfnamefont {Yong-Heng}\ \bibnamefont {Lu}}, \bibinfo
  {author} {\bibfnamefont {Jun}\ \bibnamefont {Gao}}, \bibinfo {author}
  {\bibfnamefont {Yi-Jun}\ \bibnamefont {Chang}}, \bibinfo {author}
  {\bibfnamefont {Lu-Feng}\ \bibnamefont {Qiao}}, \bibinfo {author}
  {\bibfnamefont {Zhi-Qiang}\ \bibnamefont {Jiao}}, \bibinfo {author}
  {\bibfnamefont {Hao}\ \bibnamefont {Tang}}, \ and\ \bibinfo {author}
  {\bibfnamefont {Xian-Min}\ \bibnamefont {Jin}},\ }\bibfield  {title}
  {\enquote {\bibinfo {title} {Topological protection of two-photon quantum
  correlation on a photonic chip},}\ }\href@noop {} {\bibfield  {journal}
  {\bibinfo  {journal} {Optica}\ }\textbf {\bibinfo {volume} {6}},\ \bibinfo
  {pages} {955--960} (\bibinfo {year} {2019}{\natexlab{a}})}\BibitemShut
  {NoStop}%
\bibitem [{\citenamefont {Wang}\ \emph
  {et~al.}(2019{\natexlab{b}})\citenamefont {Wang}, \citenamefont {Lu},
  \citenamefont {Mei}, \citenamefont {Gao}, \citenamefont {Li}, \citenamefont
  {Tang}, \citenamefont {Zhu}, \citenamefont {Jia},\ and\ \citenamefont
  {Jin}}]{wang2019direct}%
  \BibitemOpen
  \bibfield  {author} {\bibinfo {author} {\bibfnamefont {Yao}\ \bibnamefont
  {Wang}}, \bibinfo {author} {\bibfnamefont {Yong-Heng}\ \bibnamefont {Lu}},
  \bibinfo {author} {\bibfnamefont {Feng}\ \bibnamefont {Mei}}, \bibinfo
  {author} {\bibfnamefont {Jun}\ \bibnamefont {Gao}}, \bibinfo {author}
  {\bibfnamefont {Zhan-Ming}\ \bibnamefont {Li}}, \bibinfo {author}
  {\bibfnamefont {Hao}\ \bibnamefont {Tang}}, \bibinfo {author} {\bibfnamefont
  {Shi-Liang}\ \bibnamefont {Zhu}}, \bibinfo {author} {\bibfnamefont {Suotang}\
  \bibnamefont {Jia}}, \ and\ \bibinfo {author} {\bibfnamefont {Xian-Min}\
  \bibnamefont {Jin}},\ }\bibfield  {title} {\enquote {\bibinfo {title} {Direct
  observation of topology from single-photon dynamics},}\ }\href@noop {}
  {\bibfield  {journal} {\bibinfo  {journal} {Phys. Rev. Lett.}\ }\textbf
  {\bibinfo {volume} {122}},\ \bibinfo {pages} {193903} (\bibinfo {year}
  {2019}{\natexlab{b}})}\BibitemShut {NoStop}%
\bibitem [{\citenamefont {Barik}\ \emph {et~al.}(2018)\citenamefont {Barik},
  \citenamefont {Karasahin}, \citenamefont {Flower}, \citenamefont {Cai},
  \citenamefont {Miyake}, \citenamefont {DeGottardi}, \citenamefont {Hafezi},\
  and\ \citenamefont {Waks}}]{barik2018topological}%
  \BibitemOpen
  \bibfield  {author} {\bibinfo {author} {\bibfnamefont {Sabyasachi}\
  \bibnamefont {Barik}}, \bibinfo {author} {\bibfnamefont {Aziz}\ \bibnamefont
  {Karasahin}}, \bibinfo {author} {\bibfnamefont {Christopher}\ \bibnamefont
  {Flower}}, \bibinfo {author} {\bibfnamefont {Tao}\ \bibnamefont {Cai}},
  \bibinfo {author} {\bibfnamefont {Hirokazu}\ \bibnamefont {Miyake}}, \bibinfo
  {author} {\bibfnamefont {Wade}\ \bibnamefont {DeGottardi}}, \bibinfo {author}
  {\bibfnamefont {Mohammad}\ \bibnamefont {Hafezi}}, \ and\ \bibinfo {author}
  {\bibfnamefont {Edo}\ \bibnamefont {Waks}},\ }\bibfield  {title} {\enquote
  {\bibinfo {title} {A topological quantum optics interface},}\ }\href@noop {}
  {\bibfield  {journal} {\bibinfo  {journal} {Science}\ }\textbf {\bibinfo
  {volume} {359}},\ \bibinfo {pages} {666--668} (\bibinfo {year}
  {2018})}\BibitemShut {NoStop}%
\bibitem [{\citenamefont {Mehrabad}\ \emph {et~al.}(2019)\citenamefont
  {Mehrabad}, \citenamefont {Foster}, \citenamefont {Dost}, \citenamefont
  {Fox}, \citenamefont {Skolnick},\ and\ \citenamefont
  {Wilson}}]{mehrabad2019chiral}%
  \BibitemOpen
  \bibfield  {author} {\bibinfo {author} {\bibfnamefont {Mahmoud~Jalali}\
  \bibnamefont {Mehrabad}}, \bibinfo {author} {\bibfnamefont {Andrew~P}\
  \bibnamefont {Foster}}, \bibinfo {author} {\bibfnamefont {Ren{\'e}}\
  \bibnamefont {Dost}}, \bibinfo {author} {\bibfnamefont {A~Mark}\ \bibnamefont
  {Fox}}, \bibinfo {author} {\bibfnamefont {Maurice~S}\ \bibnamefont
  {Skolnick}}, \ and\ \bibinfo {author} {\bibfnamefont {Luke~R}\ \bibnamefont
  {Wilson}},\ }\bibfield  {title} {\enquote {\bibinfo {title} {Chiral
  topological photonics with an embedded quantum emitter},}\ }\href@noop {}
  {\bibfield  {journal} {\bibinfo  {journal} {arXiv preprint arXiv:1912.09943}\
  } (\bibinfo {year} {2019})}\BibitemShut {NoStop}%
\bibitem [{\citenamefont {Xie}\ \emph {et~al.}(2018)\citenamefont {Xie},
  \citenamefont {Wang}, \citenamefont {Wang}, \citenamefont {Zhu},
  \citenamefont {Jiang}, \citenamefont {Lu},\ and\ \citenamefont
  {Chen}}]{xie2018second}%
  \BibitemOpen
  \bibfield  {author} {\bibinfo {author} {\bibfnamefont {Bi-Ye}\ \bibnamefont
  {Xie}}, \bibinfo {author} {\bibfnamefont {Hong-Fei}\ \bibnamefont {Wang}},
  \bibinfo {author} {\bibfnamefont {Hai-Xiao}\ \bibnamefont {Wang}}, \bibinfo
  {author} {\bibfnamefont {Xue-Yi}\ \bibnamefont {Zhu}}, \bibinfo {author}
  {\bibfnamefont {Jian-Hua}\ \bibnamefont {Jiang}}, \bibinfo {author}
  {\bibfnamefont {Ming-Hui}\ \bibnamefont {Lu}}, \ and\ \bibinfo {author}
  {\bibfnamefont {Yan-Feng}\ \bibnamefont {Chen}},\ }\bibfield  {title}
  {\enquote {\bibinfo {title} {Second-order photonic topological insulator with
  corner states},}\ }\href@noop {} {\bibfield  {journal} {\bibinfo  {journal}
  {Phys. Rev. B}\ }\textbf {\bibinfo {volume} {98}},\ \bibinfo {pages} {205147}
  (\bibinfo {year} {2018})}\BibitemShut {NoStop}%
\bibitem [{\citenamefont {Xie}\ \emph {et~al.}(2019)\citenamefont {Xie},
  \citenamefont {Su}, \citenamefont {Wang}, \citenamefont {Su}, \citenamefont
  {Shen}, \citenamefont {Zhan}, \citenamefont {Lu}, \citenamefont {Wang},\ and\
  \citenamefont {Chen}}]{xie2019visualization}%
  \BibitemOpen
  \bibfield  {author} {\bibinfo {author} {\bibfnamefont {Bi-Ye}\ \bibnamefont
  {Xie}}, \bibinfo {author} {\bibfnamefont {Guang-Xu}\ \bibnamefont {Su}},
  \bibinfo {author} {\bibfnamefont {Hong-Fei}\ \bibnamefont {Wang}}, \bibinfo
  {author} {\bibfnamefont {Hai}\ \bibnamefont {Su}}, \bibinfo {author}
  {\bibfnamefont {Xiao-Peng}\ \bibnamefont {Shen}}, \bibinfo {author}
  {\bibfnamefont {Peng}\ \bibnamefont {Zhan}}, \bibinfo {author} {\bibfnamefont
  {Ming-Hui}\ \bibnamefont {Lu}}, \bibinfo {author} {\bibfnamefont {Zhen-Lin}\
  \bibnamefont {Wang}}, \ and\ \bibinfo {author} {\bibfnamefont {Yan-Feng}\
  \bibnamefont {Chen}},\ }\bibfield  {title} {\enquote {\bibinfo {title}
  {Visualization of higher-order topological insulating phases in
  two-dimensional dielectric photonic crystals},}\ }\href@noop {} {\bibfield
  {journal} {\bibinfo  {journal} {Phys. Rev. Lett.}\ }\textbf {\bibinfo
  {volume} {122}},\ \bibinfo {pages} {233903} (\bibinfo {year}
  {2019})}\BibitemShut {NoStop}%
\bibitem [{\citenamefont {Ota}\ \emph {et~al.}(2019{\natexlab{b}})\citenamefont
  {Ota}, \citenamefont {Liu}, \citenamefont {Katsumi}, \citenamefont
  {Watanabe}, \citenamefont {Wakabayashi}, \citenamefont {Arakawa},\ and\
  \citenamefont {Iwamoto}}]{ota2019photonic}%
  \BibitemOpen
  \bibfield  {author} {\bibinfo {author} {\bibfnamefont {Yasutomo}\
  \bibnamefont {Ota}}, \bibinfo {author} {\bibfnamefont {Feng}\ \bibnamefont
  {Liu}}, \bibinfo {author} {\bibfnamefont {Ryota}\ \bibnamefont {Katsumi}},
  \bibinfo {author} {\bibfnamefont {Katsuyuki}\ \bibnamefont {Watanabe}},
  \bibinfo {author} {\bibfnamefont {Katsunori}\ \bibnamefont {Wakabayashi}},
  \bibinfo {author} {\bibfnamefont {Yasuhiko}\ \bibnamefont {Arakawa}}, \ and\
  \bibinfo {author} {\bibfnamefont {Satoshi}\ \bibnamefont {Iwamoto}},\
  }\bibfield  {title} {\enquote {\bibinfo {title} {Photonic crystal nanocavity
  based on a topological corner state},}\ }\href@noop {} {\bibfield  {journal}
  {\bibinfo  {journal} {Optica}\ }\textbf {\bibinfo {volume} {6}},\ \bibinfo
  {pages} {786--789} (\bibinfo {year} {2019}{\natexlab{b}})}\BibitemShut
  {NoStop}%
\bibitem [{\citenamefont {Chen}\ \emph {et~al.}(2019)\citenamefont {Chen},
  \citenamefont {Deng}, \citenamefont {Shi}, \citenamefont {Zhao},
  \citenamefont {Chen},\ and\ \citenamefont {Dong}}]{chen2019direct}%
  \BibitemOpen
  \bibfield  {author} {\bibinfo {author} {\bibfnamefont {Xiao-Dong}\
  \bibnamefont {Chen}}, \bibinfo {author} {\bibfnamefont {Wei-Min}\
  \bibnamefont {Deng}}, \bibinfo {author} {\bibfnamefont {Fu-Long}\
  \bibnamefont {Shi}}, \bibinfo {author} {\bibfnamefont {Fu-Li}\ \bibnamefont
  {Zhao}}, \bibinfo {author} {\bibfnamefont {Min}\ \bibnamefont {Chen}}, \ and\
  \bibinfo {author} {\bibfnamefont {Jian-Wen}\ \bibnamefont {Dong}},\
  }\bibfield  {title} {\enquote {\bibinfo {title} {Direct observation of corner
  states in second-order topological photonic crystal slabs},}\ }\href@noop {}
  {\bibfield  {journal} {\bibinfo  {journal} {Phys. Rev. Lett.}\ }\textbf
  {\bibinfo {volume} {122}},\ \bibinfo {pages} {233902} (\bibinfo {year}
  {2019})}\BibitemShut {NoStop}%
\bibitem [{\citenamefont {Langbehn}\ \emph {et~al.}(2017)\citenamefont
  {Langbehn}, \citenamefont {Peng}, \citenamefont {Trifunovic}, \citenamefont
  {von Oppen},\ and\ \citenamefont {Brouwer}}]{langbehn2017reflection}%
  \BibitemOpen
  \bibfield  {author} {\bibinfo {author} {\bibfnamefont {Josias}\ \bibnamefont
  {Langbehn}}, \bibinfo {author} {\bibfnamefont {Yang}\ \bibnamefont {Peng}},
  \bibinfo {author} {\bibfnamefont {Luka}\ \bibnamefont {Trifunovic}}, \bibinfo
  {author} {\bibfnamefont {Felix}\ \bibnamefont {von Oppen}}, \ and\ \bibinfo
  {author} {\bibfnamefont {Piet~W}\ \bibnamefont {Brouwer}},\ }\bibfield
  {title} {\enquote {\bibinfo {title} {Reflection-symmetric second-order
  topological insulators and superconductors},}\ }\href@noop {} {\bibfield
  {journal} {\bibinfo  {journal} {Phys. Rev. Lett.}\ }\textbf {\bibinfo
  {volume} {119}},\ \bibinfo {pages} {246401} (\bibinfo {year}
  {2017})}\BibitemShut {NoStop}%
\bibitem [{\citenamefont {Ezawa}(2018)}]{ezawa2018higher}%
  \BibitemOpen
  \bibfield  {author} {\bibinfo {author} {\bibfnamefont {Motohiko}\
  \bibnamefont {Ezawa}},\ }\bibfield  {title} {\enquote {\bibinfo {title}
  {Higher-order topological insulators and semimetals on the breathing kagome
  and pyrochlore lattices},}\ }\href@noop {} {\bibfield  {journal} {\bibinfo
  {journal} {Phys. Rev. Lett.}\ }\textbf {\bibinfo {volume} {120}},\ \bibinfo
  {pages} {026801} (\bibinfo {year} {2018})}\BibitemShut {NoStop}%
\bibitem [{\citenamefont {Ni}\ \emph {et~al.}(2019{\natexlab{a}})\citenamefont
  {Ni}, \citenamefont {Weiner}, \citenamefont {Al{\`u}},\ and\ \citenamefont
  {Khanikaev}}]{ni2019observation}%
  \BibitemOpen
  \bibfield  {author} {\bibinfo {author} {\bibfnamefont {Xiang}\ \bibnamefont
  {Ni}}, \bibinfo {author} {\bibfnamefont {Matthew}\ \bibnamefont {Weiner}},
  \bibinfo {author} {\bibfnamefont {Andrea}\ \bibnamefont {Al{\`u}}}, \ and\
  \bibinfo {author} {\bibfnamefont {Alexander~B}\ \bibnamefont {Khanikaev}},\
  }\bibfield  {title} {\enquote {\bibinfo {title} {Observation of higher-order
  topological acoustic states protected by generalized chiral symmetry},}\
  }\href@noop {} {\bibfield  {journal} {\bibinfo  {journal} {Nat. Mater.}\
  }\textbf {\bibinfo {volume} {18}},\ \bibinfo {pages} {113} (\bibinfo {year}
  {2019}{\natexlab{a}})}\BibitemShut {NoStop}%
\bibitem [{\citenamefont {Noh}\ \emph {et~al.}(2018)\citenamefont {Noh},
  \citenamefont {Benalcazar}, \citenamefont {Huang}, \citenamefont {Collins},
  \citenamefont {Chen}, \citenamefont {Hughes},\ and\ \citenamefont
  {Rechtsman}}]{noh2018topological}%
  \BibitemOpen
  \bibfield  {author} {\bibinfo {author} {\bibfnamefont {Jiho}\ \bibnamefont
  {Noh}}, \bibinfo {author} {\bibfnamefont {Wladimir~A}\ \bibnamefont
  {Benalcazar}}, \bibinfo {author} {\bibfnamefont {Sheng}\ \bibnamefont
  {Huang}}, \bibinfo {author} {\bibfnamefont {Matthew~J}\ \bibnamefont
  {Collins}}, \bibinfo {author} {\bibfnamefont {Kevin~P}\ \bibnamefont {Chen}},
  \bibinfo {author} {\bibfnamefont {Taylor~L}\ \bibnamefont {Hughes}}, \ and\
  \bibinfo {author} {\bibfnamefont {Mikael~C}\ \bibnamefont {Rechtsman}},\
  }\bibfield  {title} {\enquote {\bibinfo {title} {Topological protection of
  photonic mid-gap defect modes},}\ }\href@noop {} {\bibfield  {journal}
  {\bibinfo  {journal} {Nat. Photonics}\ }\textbf {\bibinfo {volume} {12}},\
  \bibinfo {pages} {408} (\bibinfo {year} {2018})}\BibitemShut {NoStop}%
\bibitem [{\citenamefont {Benalcazar}\ \emph
  {et~al.}(2017{\natexlab{a}})\citenamefont {Benalcazar}, \citenamefont
  {Bernevig},\ and\ \citenamefont {Hughes}}]{benalcazar2017quantized}%
  \BibitemOpen
  \bibfield  {author} {\bibinfo {author} {\bibfnamefont {Wladimir~A}\
  \bibnamefont {Benalcazar}}, \bibinfo {author} {\bibfnamefont {B~Andrei}\
  \bibnamefont {Bernevig}}, \ and\ \bibinfo {author} {\bibfnamefont {Taylor~L}\
  \bibnamefont {Hughes}},\ }\bibfield  {title} {\enquote {\bibinfo {title}
  {Quantized electric multipole insulators},}\ }\href@noop {} {\bibfield
  {journal} {\bibinfo  {journal} {Science}\ }\textbf {\bibinfo {volume}
  {357}},\ \bibinfo {pages} {61--66} (\bibinfo {year}
  {2017}{\natexlab{a}})}\BibitemShut {NoStop}%
\bibitem [{\citenamefont {Benalcazar}\ \emph
  {et~al.}(2017{\natexlab{b}})\citenamefont {Benalcazar}, \citenamefont
  {Bernevig},\ and\ \citenamefont {Hughes}}]{benalcazar2017electric}%
  \BibitemOpen
  \bibfield  {author} {\bibinfo {author} {\bibfnamefont {Wladimir~A}\
  \bibnamefont {Benalcazar}}, \bibinfo {author} {\bibfnamefont {B~Andrei}\
  \bibnamefont {Bernevig}}, \ and\ \bibinfo {author} {\bibfnamefont {Taylor~L}\
  \bibnamefont {Hughes}},\ }\bibfield  {title} {\enquote {\bibinfo {title}
  {Electric multipole moments, topological multipole moment pumping, and chiral
  hinge states in crystalline insulators},}\ }\href@noop {} {\bibfield
  {journal} {\bibinfo  {journal} {Phys. Rev. B}\ }\textbf {\bibinfo {volume}
  {96}},\ \bibinfo {pages} {245115} (\bibinfo {year}
  {2017}{\natexlab{b}})}\BibitemShut {NoStop}%
\bibitem [{\citenamefont {Imhof}\ \emph {et~al.}(2018)\citenamefont {Imhof},
  \citenamefont {Berger}, \citenamefont {Bayer}, \citenamefont {Brehm},
  \citenamefont {Molenkamp}, \citenamefont {Kiessling}, \citenamefont
  {Schindler}, \citenamefont {Lee}, \citenamefont {Greiter}, \citenamefont
  {Neupert} \emph {et~al.}}]{imhof2018topolectrical}%
  \BibitemOpen
  \bibfield  {author} {\bibinfo {author} {\bibfnamefont {Stefan}\ \bibnamefont
  {Imhof}}, \bibinfo {author} {\bibfnamefont {Christian}\ \bibnamefont
  {Berger}}, \bibinfo {author} {\bibfnamefont {Florian}\ \bibnamefont {Bayer}},
  \bibinfo {author} {\bibfnamefont {Johannes}\ \bibnamefont {Brehm}}, \bibinfo
  {author} {\bibfnamefont {Laurens~W}\ \bibnamefont {Molenkamp}}, \bibinfo
  {author} {\bibfnamefont {Tobias}\ \bibnamefont {Kiessling}}, \bibinfo
  {author} {\bibfnamefont {Frank}\ \bibnamefont {Schindler}}, \bibinfo {author}
  {\bibfnamefont {Ching~Hua}\ \bibnamefont {Lee}}, \bibinfo {author}
  {\bibfnamefont {Martin}\ \bibnamefont {Greiter}}, \bibinfo {author}
  {\bibfnamefont {Titus}\ \bibnamefont {Neupert}},  \emph {et~al.},\ }\bibfield
   {title} {\enquote {\bibinfo {title} {Topolectrical-circuit realization of
  topological corner modes},}\ }\href@noop {} {\bibfield  {journal} {\bibinfo
  {journal} {Nat. Phys.}\ }\textbf {\bibinfo {volume} {14}},\ \bibinfo {pages}
  {925} (\bibinfo {year} {2018})}\BibitemShut {NoStop}%
\bibitem [{\citenamefont {Peterson}\ \emph {et~al.}(2018)\citenamefont
  {Peterson}, \citenamefont {Benalcazar}, \citenamefont {Hughes},\ and\
  \citenamefont {Bahl}}]{peterson2018quantized}%
  \BibitemOpen
  \bibfield  {author} {\bibinfo {author} {\bibfnamefont {Christopher~W}\
  \bibnamefont {Peterson}}, \bibinfo {author} {\bibfnamefont {Wladimir~A}\
  \bibnamefont {Benalcazar}}, \bibinfo {author} {\bibfnamefont {Taylor~L}\
  \bibnamefont {Hughes}}, \ and\ \bibinfo {author} {\bibfnamefont {Gaurav}\
  \bibnamefont {Bahl}},\ }\bibfield  {title} {\enquote {\bibinfo {title} {A
  quantized microwave quadrupole insulator with topologically protected corner
  states},}\ }\href@noop {} {\bibfield  {journal} {\bibinfo  {journal}
  {Nature}\ }\textbf {\bibinfo {volume} {555}},\ \bibinfo {pages} {346}
  (\bibinfo {year} {2018})}\BibitemShut {NoStop}%
\bibitem [{\citenamefont {Serra-Garcia}\ \emph {et~al.}(2018)\citenamefont
  {Serra-Garcia}, \citenamefont {Peri}, \citenamefont {S{\"u}sstrunk},
  \citenamefont {Bilal}, \citenamefont {Larsen}, \citenamefont {Villanueva},\
  and\ \citenamefont {Huber}}]{serra2018observation}%
  \BibitemOpen
  \bibfield  {author} {\bibinfo {author} {\bibfnamefont {Marc}\ \bibnamefont
  {Serra-Garcia}}, \bibinfo {author} {\bibfnamefont {Valerio}\ \bibnamefont
  {Peri}}, \bibinfo {author} {\bibfnamefont {Roman}\ \bibnamefont
  {S{\"u}sstrunk}}, \bibinfo {author} {\bibfnamefont {Osama~R}\ \bibnamefont
  {Bilal}}, \bibinfo {author} {\bibfnamefont {Tom}\ \bibnamefont {Larsen}},
  \bibinfo {author} {\bibfnamefont {Luis~Guillermo}\ \bibnamefont
  {Villanueva}}, \ and\ \bibinfo {author} {\bibfnamefont {Sebastian~D}\
  \bibnamefont {Huber}},\ }\bibfield  {title} {\enquote {\bibinfo {title}
  {Observation of a phononic quadrupole topological insulator},}\ }\href@noop
  {} {\bibfield  {journal} {\bibinfo  {journal} {Nature}\ }\textbf {\bibinfo
  {volume} {555}},\ \bibinfo {pages} {342} (\bibinfo {year}
  {2018})}\BibitemShut {NoStop}%
\bibitem [{\citenamefont {Mittal}\ \emph {et~al.}(2019)\citenamefont {Mittal},
  \citenamefont {Orre}, \citenamefont {Zhu}, \citenamefont {Gorlach},
  \citenamefont {Poddubny},\ and\ \citenamefont {Hafezi}}]{mittal2019photonic}%
  \BibitemOpen
  \bibfield  {author} {\bibinfo {author} {\bibfnamefont {Sunil}\ \bibnamefont
  {Mittal}}, \bibinfo {author} {\bibfnamefont {Venkata~Vikram}\ \bibnamefont
  {Orre}}, \bibinfo {author} {\bibfnamefont {Guanyu}\ \bibnamefont {Zhu}},
  \bibinfo {author} {\bibfnamefont {Maxim~A}\ \bibnamefont {Gorlach}}, \bibinfo
  {author} {\bibfnamefont {Alexander}\ \bibnamefont {Poddubny}}, \ and\
  \bibinfo {author} {\bibfnamefont {Mohammad}\ \bibnamefont {Hafezi}},\
  }\bibfield  {title} {\enquote {\bibinfo {title} {Photonic quadrupole
  topological phases},}\ }\href@noop {} {\bibfield  {journal} {\bibinfo
  {journal} {Nature Photonics}\ }\textbf {\bibinfo {volume} {13}},\ \bibinfo
  {pages} {692--696} (\bibinfo {year} {2019})}\BibitemShut {NoStop}%
\bibitem [{\citenamefont {Xue}\ \emph {et~al.}(2019{\natexlab{a}})\citenamefont
  {Xue}, \citenamefont {Yang}, \citenamefont {Gao}, \citenamefont {Chong},\
  and\ \citenamefont {Zhang}}]{xue2019acoustic}%
  \BibitemOpen
  \bibfield  {author} {\bibinfo {author} {\bibfnamefont {Haoran}\ \bibnamefont
  {Xue}}, \bibinfo {author} {\bibfnamefont {Yahui}\ \bibnamefont {Yang}},
  \bibinfo {author} {\bibfnamefont {Fei}\ \bibnamefont {Gao}}, \bibinfo
  {author} {\bibfnamefont {Yidong}\ \bibnamefont {Chong}}, \ and\ \bibinfo
  {author} {\bibfnamefont {Baile}\ \bibnamefont {Zhang}},\ }\bibfield  {title}
  {\enquote {\bibinfo {title} {Acoustic higher-order topological insulator on a
  kagome lattice},}\ }\href@noop {} {\bibfield  {journal} {\bibinfo  {journal}
  {Nat. Mater.}\ }\textbf {\bibinfo {volume} {18}},\ \bibinfo {pages} {108}
  (\bibinfo {year} {2019}{\natexlab{a}})}\BibitemShut {NoStop}%
\bibitem [{\citenamefont {Zhang}\ \emph {et~al.}(2019)\citenamefont {Zhang},
  \citenamefont {Wang}, \citenamefont {Lin}, \citenamefont {Tian},
  \citenamefont {Xie}, \citenamefont {Lu}, \citenamefont {Chen},\ and\
  \citenamefont {Jiang}}]{zhang2019second}%
  \BibitemOpen
  \bibfield  {author} {\bibinfo {author} {\bibfnamefont {Xiujuan}\ \bibnamefont
  {Zhang}}, \bibinfo {author} {\bibfnamefont {Hai-Xiao}\ \bibnamefont {Wang}},
  \bibinfo {author} {\bibfnamefont {Zhi-Kang}\ \bibnamefont {Lin}}, \bibinfo
  {author} {\bibfnamefont {Yuan}\ \bibnamefont {Tian}}, \bibinfo {author}
  {\bibfnamefont {Biye}\ \bibnamefont {Xie}}, \bibinfo {author} {\bibfnamefont
  {Ming-Hui}\ \bibnamefont {Lu}}, \bibinfo {author} {\bibfnamefont {Yan-Feng}\
  \bibnamefont {Chen}}, \ and\ \bibinfo {author} {\bibfnamefont {Jian-Hua}\
  \bibnamefont {Jiang}},\ }\bibfield  {title} {\enquote {\bibinfo {title}
  {Second-order topology and multidimensional topological transitions in sonic
  crystals},}\ }\href@noop {} {\bibfield  {journal} {\bibinfo  {journal} {Nat.
  Phys.}\ }\textbf {\bibinfo {volume} {15}},\ \bibinfo {pages} {582} (\bibinfo
  {year} {2019})}\BibitemShut {NoStop}%
\bibitem [{\citenamefont {Schindler}\ \emph {et~al.}(2018)\citenamefont
  {Schindler}, \citenamefont {Wang}, \citenamefont {Vergniory}, \citenamefont
  {Cook}, \citenamefont {Murani}, \citenamefont {Sengupta}, \citenamefont
  {Kasumov}, \citenamefont {Deblock}, \citenamefont {Jeon}, \citenamefont
  {Drozdov} \emph {et~al.}}]{schindler2018higher}%
  \BibitemOpen
  \bibfield  {author} {\bibinfo {author} {\bibfnamefont {Frank}\ \bibnamefont
  {Schindler}}, \bibinfo {author} {\bibfnamefont {Zhijun}\ \bibnamefont
  {Wang}}, \bibinfo {author} {\bibfnamefont {Maia~G}\ \bibnamefont
  {Vergniory}}, \bibinfo {author} {\bibfnamefont {Ashley~M}\ \bibnamefont
  {Cook}}, \bibinfo {author} {\bibfnamefont {Anil}\ \bibnamefont {Murani}},
  \bibinfo {author} {\bibfnamefont {Shamashis}\ \bibnamefont {Sengupta}},
  \bibinfo {author} {\bibfnamefont {Alik~Yu}\ \bibnamefont {Kasumov}}, \bibinfo
  {author} {\bibfnamefont {Richard}\ \bibnamefont {Deblock}}, \bibinfo {author}
  {\bibfnamefont {Sangjun}\ \bibnamefont {Jeon}}, \bibinfo {author}
  {\bibfnamefont {Ilya}\ \bibnamefont {Drozdov}},  \emph {et~al.},\ }\bibfield
  {title} {\enquote {\bibinfo {title} {Higher-order topology in bismuth},}\
  }\href@noop {} {\bibfield  {journal} {\bibinfo  {journal} {Nat. Phys.}\
  }\textbf {\bibinfo {volume} {14}},\ \bibinfo {pages} {918} (\bibinfo {year}
  {2018})}\BibitemShut {NoStop}%
\bibitem [{\citenamefont {El~Hassan}\ \emph {et~al.}(2019)\citenamefont
  {El~Hassan}, \citenamefont {Kunst}, \citenamefont {Moritz}, \citenamefont
  {Andler}, \citenamefont {Bergholtz},\ and\ \citenamefont
  {Bourennane}}]{el2019corner}%
  \BibitemOpen
  \bibfield  {author} {\bibinfo {author} {\bibfnamefont {Ashraf}\ \bibnamefont
  {El~Hassan}}, \bibinfo {author} {\bibfnamefont {Flore~K}\ \bibnamefont
  {Kunst}}, \bibinfo {author} {\bibfnamefont {Alexander}\ \bibnamefont
  {Moritz}}, \bibinfo {author} {\bibfnamefont {Guillermo}\ \bibnamefont
  {Andler}}, \bibinfo {author} {\bibfnamefont {Emil~J}\ \bibnamefont
  {Bergholtz}}, \ and\ \bibinfo {author} {\bibfnamefont {Mohamed}\ \bibnamefont
  {Bourennane}},\ }\bibfield  {title} {\enquote {\bibinfo {title} {Corner
  states of light in photonic waveguides},}\ }\href@noop {} {\bibfield
  {journal} {\bibinfo  {journal} {Nature Photonics}\ }\textbf {\bibinfo
  {volume} {13}},\ \bibinfo {pages} {697--700} (\bibinfo {year}
  {2019})}\BibitemShut {NoStop}%
\bibitem [{\citenamefont {Zhang}\ \emph
  {et~al.}(2020{\natexlab{a}})\citenamefont {Zhang}, \citenamefont {Zou},
  \citenamefont {He}, \citenamefont {Bao}, \citenamefont {Pei}, \citenamefont
  {Sun},\ and\ \citenamefont {Zhang}}]{zhang2020topolectrical}%
  \BibitemOpen
  \bibfield  {author} {\bibinfo {author} {\bibfnamefont {Weixuan}\ \bibnamefont
  {Zhang}}, \bibinfo {author} {\bibfnamefont {Deyuan}\ \bibnamefont {Zou}},
  \bibinfo {author} {\bibfnamefont {Wenjing}\ \bibnamefont {He}}, \bibinfo
  {author} {\bibfnamefont {Jiacheng}\ \bibnamefont {Bao}}, \bibinfo {author}
  {\bibfnamefont {Qingsong}\ \bibnamefont {Pei}}, \bibinfo {author}
  {\bibfnamefont {Houjun}\ \bibnamefont {Sun}}, \ and\ \bibinfo {author}
  {\bibfnamefont {Xiangdong}\ \bibnamefont {Zhang}},\ }\bibfield  {title}
  {\enquote {\bibinfo {title} {Topolectrical-circuit realization of 4d
  hexadecapole insulator},}\ }\href@noop {} {\bibfield  {journal} {\bibinfo
  {journal} {arXiv preprint arXiv:2001.07931}\ } (\bibinfo {year}
  {2020}{\natexlab{a}})}\BibitemShut {NoStop}%
\bibitem [{\citenamefont {Bao}\ \emph {et~al.}(2019)\citenamefont {Bao},
  \citenamefont {Zou}, \citenamefont {Zhang}, \citenamefont {He}, \citenamefont
  {Sun},\ and\ \citenamefont {Zhang}}]{bao2019topoelectrical}%
  \BibitemOpen
  \bibfield  {author} {\bibinfo {author} {\bibfnamefont {Jiacheng}\
  \bibnamefont {Bao}}, \bibinfo {author} {\bibfnamefont {Deyuan}\ \bibnamefont
  {Zou}}, \bibinfo {author} {\bibfnamefont {Weixuan}\ \bibnamefont {Zhang}},
  \bibinfo {author} {\bibfnamefont {Wenjing}\ \bibnamefont {He}}, \bibinfo
  {author} {\bibfnamefont {Houjun}\ \bibnamefont {Sun}}, \ and\ \bibinfo
  {author} {\bibfnamefont {Xiangdong}\ \bibnamefont {Zhang}},\ }\bibfield
  {title} {\enquote {\bibinfo {title} {Topoelectrical circuit octupole
  insulator with topologically protected corner states},}\ }\href@noop {}
  {\bibfield  {journal} {\bibinfo  {journal} {Phys. Rev. B}\ }\textbf {\bibinfo
  {volume} {100}},\ \bibinfo {pages} {201406} (\bibinfo {year}
  {2019})}\BibitemShut {NoStop}%
\bibitem [{\citenamefont {Xue}\ \emph {et~al.}(2019{\natexlab{b}})\citenamefont
  {Xue}, \citenamefont {Ge}, \citenamefont {Sun}, \citenamefont {Wang},
  \citenamefont {Jia}, \citenamefont {Guan}, \citenamefont {Yuan},
  \citenamefont {Chong},\ and\ \citenamefont {Zhang}}]{xue2019quantized}%
  \BibitemOpen
  \bibfield  {author} {\bibinfo {author} {\bibfnamefont {Haoran}\ \bibnamefont
  {Xue}}, \bibinfo {author} {\bibfnamefont {Yong}\ \bibnamefont {Ge}}, \bibinfo
  {author} {\bibfnamefont {Hong-Xiang}\ \bibnamefont {Sun}}, \bibinfo {author}
  {\bibfnamefont {Qiang}\ \bibnamefont {Wang}}, \bibinfo {author}
  {\bibfnamefont {Ding}\ \bibnamefont {Jia}}, \bibinfo {author} {\bibfnamefont
  {Yi-Jun}\ \bibnamefont {Guan}}, \bibinfo {author} {\bibfnamefont {Shou-Qi}\
  \bibnamefont {Yuan}}, \bibinfo {author} {\bibfnamefont {Yidong}\ \bibnamefont
  {Chong}}, \ and\ \bibinfo {author} {\bibfnamefont {Baile}\ \bibnamefont
  {Zhang}},\ }\bibfield  {title} {\enquote {\bibinfo {title} {Quantized
  octupole acoustic topological insulator},}\ }\href@noop {} {\bibfield
  {journal} {\bibinfo  {journal} {arXiv preprint arXiv:1911.06068}\ } (\bibinfo
  {year} {2019}{\natexlab{b}})}\BibitemShut {NoStop}%
\bibitem [{\citenamefont {Ni}\ \emph {et~al.}(2019{\natexlab{b}})\citenamefont
  {Ni}, \citenamefont {Li}, \citenamefont {Weiner}, \citenamefont {Al{\`u}},\
  and\ \citenamefont {Khanikaev}}]{ni2019demonstration}%
  \BibitemOpen
  \bibfield  {author} {\bibinfo {author} {\bibfnamefont {Xiang}\ \bibnamefont
  {Ni}}, \bibinfo {author} {\bibfnamefont {Mengyao}\ \bibnamefont {Li}},
  \bibinfo {author} {\bibfnamefont {Matthew}\ \bibnamefont {Weiner}}, \bibinfo
  {author} {\bibfnamefont {Andrea}\ \bibnamefont {Al{\`u}}}, \ and\ \bibinfo
  {author} {\bibfnamefont {Alexander~B}\ \bibnamefont {Khanikaev}},\ }\bibfield
   {title} {\enquote {\bibinfo {title} {Demonstration of a quantized acoustic
  octupole topological insulator},}\ }\href@noop {} {\bibfield  {journal}
  {\bibinfo  {journal} {arXiv preprint arXiv:1911.06469}\ } (\bibinfo {year}
  {2019}{\natexlab{b}})}\BibitemShut {NoStop}%
\bibitem [{\citenamefont {Zhang}\ \emph
  {et~al.}(2020{\natexlab{b}})\citenamefont {Zhang}, \citenamefont {Lin},
  \citenamefont {Wang}, \citenamefont {Xiong}, \citenamefont {Tian},
  \citenamefont {Lu}, \citenamefont {Chen},\ and\ \citenamefont
  {Jiang}}]{zhang2020symmetry}%
  \BibitemOpen
  \bibfield  {author} {\bibinfo {author} {\bibfnamefont {Xiujuan}\ \bibnamefont
  {Zhang}}, \bibinfo {author} {\bibfnamefont {Zhi-Kang}\ \bibnamefont {Lin}},
  \bibinfo {author} {\bibfnamefont {Hai-Xiao}\ \bibnamefont {Wang}}, \bibinfo
  {author} {\bibfnamefont {Zhan}\ \bibnamefont {Xiong}}, \bibinfo {author}
  {\bibfnamefont {Yuan}\ \bibnamefont {Tian}}, \bibinfo {author} {\bibfnamefont
  {Ming-Hui}\ \bibnamefont {Lu}}, \bibinfo {author} {\bibfnamefont {Yan-Feng}\
  \bibnamefont {Chen}}, \ and\ \bibinfo {author} {\bibfnamefont {Jian-Hua}\
  \bibnamefont {Jiang}},\ }\bibfield  {title} {\enquote {\bibinfo {title}
  {Symmetry-protected hierarchy of anomalous multipole topological band gaps in
  nonsymmorphic metacrystals},}\ }\href@noop {} {\bibfield  {journal} {\bibinfo
   {journal} {Nat. Commun.}\ }\textbf {\bibinfo {volume} {11}},\ \bibinfo
  {pages} {1--9} (\bibinfo {year} {2020}{\natexlab{b}})}\BibitemShut {NoStop}%
\bibitem [{\citenamefont {Zak}(1989)}]{zak1989berry}%
  \BibitemOpen
  \bibfield  {author} {\bibinfo {author} {\bibfnamefont {J}~\bibnamefont
  {Zak}},\ }\bibfield  {title} {\enquote {\bibinfo {title} {Berry’s phase for
  energy bands in solids},}\ }\href@noop {} {\bibfield  {journal} {\bibinfo
  {journal} {Phys. Rev. Lett.}\ }\textbf {\bibinfo {volume} {62}},\ \bibinfo
  {pages} {2747} (\bibinfo {year} {1989})}\BibitemShut {NoStop}%
\bibitem [{\citenamefont {Hohenester}\ \emph {et~al.}(2009)\citenamefont
  {Hohenester}, \citenamefont {Laucht}, \citenamefont {Kaniber}, \citenamefont
  {Hauke}, \citenamefont {Neumann}, \citenamefont {Mohtashami}, \citenamefont
  {Seliger}, \citenamefont {Bichler},\ and\ \citenamefont
  {Finley}}]{Hohenester2009}%
  \BibitemOpen
  \bibfield  {author} {\bibinfo {author} {\bibfnamefont {Ulrich}\ \bibnamefont
  {Hohenester}}, \bibinfo {author} {\bibfnamefont {Arne}\ \bibnamefont
  {Laucht}}, \bibinfo {author} {\bibfnamefont {Michael}\ \bibnamefont
  {Kaniber}}, \bibinfo {author} {\bibfnamefont {Norman}\ \bibnamefont {Hauke}},
  \bibinfo {author} {\bibfnamefont {Andre}\ \bibnamefont {Neumann}}, \bibinfo
  {author} {\bibfnamefont {Abbas}\ \bibnamefont {Mohtashami}}, \bibinfo
  {author} {\bibfnamefont {Marek}\ \bibnamefont {Seliger}}, \bibinfo {author}
  {\bibfnamefont {Max}\ \bibnamefont {Bichler}}, \ and\ \bibinfo {author}
  {\bibfnamefont {Jonathan~J.}\ \bibnamefont {Finley}},\ }\bibfield  {title}
  {\enquote {\bibinfo {title} {Phonon-assisted transitions from quantum dot
  excitons to cavity photons},}\ }\href@noop {} {\bibfield  {journal} {\bibinfo
   {journal} {Phys. Rev. B}\ }\textbf {\bibinfo {volume} {80}},\ \bibinfo
  {pages} {201311} (\bibinfo {year} {2009})}\BibitemShut {NoStop}%
\end{thebibliography}
\end{document}